\documentclass[sn-mathphys]{sn-jnl}

\usepackage{amsmath,amssymb,amsthm}
\usepackage{graphicx}
\usepackage{bm}
\usepackage{booktabs}
\usepackage{xcolor}
\usepackage{listings}
\usepackage{microtype}
\usepackage{enumitem}
\usepackage{tcolorbox}

\newcommand{\vqe}{\text{VQE}}

\newcommand{\KO}{\tilde{K}_0}
\newcommand{\DS}{D_S}

\begin{document}

\title[Quantum Computing Solution of the Bethe-Salpeter Equation]{Quantum Computing Solution of the Bethe-Salpeter Equation for Relativistic Scalar Bound States via Tensor-Network VQE}

\author*[1]{\fnm{Gerhard} \sur{Hellstern}}
\email{gerhard.hellstern@dhbw-stuttgart.de}

\affil[1]{\orgdiv{Duale Hochschule Baden-W\"urttemberg Stuttgart, Herdweg 20, 70174 Stuttgart}, 
\country{Germany}}

\abstract{We present a gate-based quantum computing solution of the homogeneous 
Bethe-Salpeter equation (hBSE) for the bound state of two massive relativistic 
scalar particles interacting via ladder-approximation scalar exchange. 
After Wick rotation to Euclidean space and O(4) S-wave partial-wave projection, 
the hBSE is reduced to a symmetric matrix eigenvalue problem of dimension 
$N = 2^n$. We decompose the resulting Hamiltonian into a sum of $n$-qubit Pauli 
operators and solve it with the Variational Quantum Eigensolver (VQE) using 
a Matrix Product State (MPS) tensor-network ansatz. For $N=16$ ($n=4$ qubits), 
the VQE recovers the maximum eigenvalue—which encodes the minimum coupling 
constant for binding—to better than $1\%$ mean relative error compared to 
classical diagonalization. Entanglement analysis of the BSE amplitude shows low, 
area-law-like entanglement over the tested sizes, which 
 motivates the MPS ansatz. A critical analysis of three independent barriers - 
 exponential Pauli overhead, approximately size-independent low entanglement, 
 and decreasing VQE gradient scales for the tested ansatz, consistent with 
 generic barren-plateau concerns at larger $n$ -  reveals that the specific 
 problem studied here lies in a classically tractable regime: over the tested
 range it is well described by low-bond-dimension tensor networks and 
  efficiently handled by MPS/Lanczos methods. This negative result provides, 
  to our knowledge, the first entanglement quantification of the BSE amplitude 
  in qubit encoding, establishes the Pauli Hamiltonian framework for future BSE 
  variants, and identifies 2D Minkowski-space BSE, $N$-body bound states, and 
  non-ladder kernels as physically motivated extensions where genuine quantum 
  advantage may become plausible.}


\maketitle

\section{\label{sec:intro}Introduction}

The Bethe-Salpeter equation (BSE)~\cite{Bethe1951} is the
covariant, non-perturbative description of relativistic two-body bound states
in quantum field theory (QFT). Since its formulation in 1951, it has been applied
to positronium and hydrogen in QED~\cite{SchwartzZemach1966,Kutzelnigg2022},
mesons and baryons in QCD~\cite{Roberts1994,Maris1997,Hellstern1997,Eichmann2016,Barabanov2021,Ding2024}, including
covariant quark-diquark treatments of the nucleon in ladder approximation~\cite{Hellstern1997},
and to excitons in condensed matter~\cite{HankeSham1977}. Despite its central importance,
numerical solutions of the BSE remain computationally expensive, particularly
when moving beyond the ladder approximation or working directly in Minkowski
space~\cite{Karmanov2006,Carbonell2010,Frederico2012,Frederico2014,Fornetti2024annealer}.

Quantum computing (QC) offers a fundamentally different approach to eigenvalue
problems in QFT. Variational Quantum Eigensolvers
(VQE)~\cite{Peruzzo2014,McClean2016,Tilly2022} can encode exponentially large
Hilbert spaces in polynomial qubit registers, and tensor-network
backends~\cite{Schollwoeck2011,Orus2014} further reduce circuit costs by
exploiting low-entanglement structure when it is present in the target
wavefunction.

Recent years have seen the first quantum simulation of QFT bound states: light-front
quantization was combined with VQE to compute pion structure on IBM quantum
hardware~\cite{Kreshchuk2021,Kreshchuk2021b,Liu2021}, and the BSE itself was
solved for the first time on a D-Wave quantum annealer via a Quadratic Unconstrained
Binary Optimization (QUBO) reformulation~\cite{Fornetti2024annealer}. However, the
gate-based VQE approach—which provides a natural pathway to fault-tolerant
quantum algorithms and scales differently with system size—has not yet been
applied to the covariant BSE, to our knowledge.

In this paper, we address this gap. We solve the homogeneous BSE (hBSE) for two
identical massive scalars exchanging a massive scalar in the ladder approximation,
using the following strategy:
\begin{enumerate}
  \item Wick-rotate to Euclidean space and project to the O(4) S-wave sector,
        reducing the 4D integral equation to a 1D radial equation.
  \item Discretize with Gauss-Legendre quadrature, yielding a symmetric
        $N\times N$ matrix eigenvalue problem.
  \item Decompose the Hamiltonian into $n$-qubit Pauli strings and apply VQE
        with an MPS tensor-network ansatz.
    \item Benchmark against the classical eigenvalue calculation and analyze
      the entanglement structure over accessible sizes to motivate the MPS ansatz.
\end{enumerate}

The paper is organized as follows. Section~\ref{sec:bse} reviews the BSE and its
Euclidean reduction. Section~\ref{sec:discretization} presents the discretization
and symmetrization into a standard eigenvalue problem. Section~\ref{sec:quantum}
describes the quantum computing formulation: Pauli decomposition, MPS ansatz, and
VQE. Section~\ref{sec:results} presents the benchmark results at $N=16$,
Section~\ref{sec:largeN} extends the analysis to larger discretizations, and
Section~\ref{sec:outlook} discusses complexity, quantum advantage, and future
directions.

\section{\label{sec:bse}The Bethe-Salpeter Equation}

\subsection{General formulation}

Physically, the homogeneous BSE is the relativistic, field-theoretic analogue
of the bound-state Schr\"odinger equation: it resums the infinite ladder of
exchanged particles of mass $\mu$ that binds two constituent particles of mass $m$, 
and admits non-trivial solutions
only for discrete pairs of coupling and bound-state mass, just as the
Schr\"odinger equation admits bound states only at discrete energies. The
Bethe-Salpeter amplitude $\Gamma(P,p)$ plays the role of the bound-state wave
function in momentum space. Here, we focus on the simplest non-trivial case of 
two identical scalars interacting via scalar exchange.

The homogeneous BSE for the two-body Bethe-Salpeter amplitude $\Gamma(P,p)$
in Minkowski space reads~\cite{Bethe1951}:
\begin{equation}
  \Gamma(P,p) = \mathrm{i}\!\int\!\frac{d^4k}{(2\pi)^4}\,
  K(P,p,k)\,S_1\!\left(\tfrac{P}{2}+k\right)
  S_2\!\left(\tfrac{P}{2}-k\right)\Gamma(P,k),
  \label{eq:bse_full}
\end{equation}
where $P$ is the total 4-momentum, $p$ is the relative 4-momentum,
$K$ is the two-particle-irreducible kernel, and $S_i$ are the full
propagators of the constituent particles. Here $g$ is the coupling constant of the
constituent--exchange vertex, $\mu$ the mass of the exchanged scalar particle, $m$ the
mass of each constituent, and $M$ the mass of the bound state, restricted to
$M<2m$ so that the state lies below the two-particle threshold.  The amplitude
$\Gamma(P,p)$ is the momentum-space bound-state wave function; its O(4) S-wave
projection $\phi(p)$, introduced in Sec.~\ref{sec:bse} below, is the object
actually discretised and solved for.

For two identical scalar particles of mass $m$ interacting via exchange of
a scalar of mass $\mu$, in the ladder approximation, the kernel reduces to:
\begin{equation}
  K(P,p,k) = \frac{-\mathrm{i}\,g^2}{(p-k)^2 - \mu^2 + \mathrm{i}\epsilon}.
  \label{eq:kernel_ladder}
\end{equation}
The free scalar propagator is $S(k) = \mathrm{i}/(k^2-m^2+\mathrm{i}\epsilon)$
and 
$\mathrm{i}\epsilon$ is the usual Feynman pole prescription.

\subsection{Wick rotation and Euclidean reduction}

We perform a Wick rotation $p^0 \to \mathrm{i}p^4_E$ to pass to Euclidean
space~\cite{Wick1954}. In the rest frame $P^\mu = (M,\bm{0})$, with
$M < 2m$ for a bound state, the constituent particle Euclidean momenta become:
\begin{align}
  k_{1,E} &= P_E/2 + p_E, \quad
  k_{2,E}  = P_E/2 - p_E,
\end{align}
with $P_E = (0,0,0,M)$. The inverse Euclidean propagator product is:
\begin{equation}
  D(p_E) = \bigl(k_{1,E}^2 + m^2\bigr)\bigl(k_{2,E}^2 + m^2\bigr).
\end{equation}
The radial (single-channel) reduction below requires the O(4) angular average of
this product over the unit sphere $S^3$. We therefore define the
\emph{O(4)-projected inverse propagator product}
\begin{equation}
  \DS(p,M) \;\equiv\;
  \bigl\langle (k_{1E}^2+m^2)(k_{2E}^2+m^2) \bigr\rangle_{S^3},
  \label{eq:DS_def}
\end{equation}
with $p\equiv|p_E|$. Its closed form is evaluated in
Eqs.~\eqref{eq:D_avg}--\eqref{eq:D_propagator} below; $\DS$ thus replaces the
angle-dependent product $D(p_E)$ by its $S^3$ average, which is the only
approximation made in passing to the single S-wave channel.

\subsection{O(4) partial-wave projection}

After the Wick rotation the rest-frame kernel is invariant under O(4) rotations
of the Euclidean relative momentum $p_E$, so the amplitude can be expanded in
O(4) hyperspherical harmonics. Retaining only the lowest (S-wave, $\ell=0$)
component defines the radial amplitude as the angular average of the
Wick-rotated amplitude $\Gamma_E$ over $S^3$,
\begin{equation}
  \phi(p) \;\equiv\; \bigl\langle \Gamma_E(P_E,p_E) \bigr\rangle_{S^3}
  \;=\; \frac{1}{\Omega_4}\int_{S^3}\! d\Omega_3\,\Gamma_E(P_E,p_E),
  \qquad p\equiv|p_E|,
  \label{eq:phi_def}
\end{equation}
which is the bound-state wave function used from here on. Here $d\Omega_3$ is the
surface element on the unit sphere $S^3$ and
$\Omega_4 \equiv \int_{S^3} d\Omega_3 = 2\pi^2$ is the total O(4) solid angle.
Higher partial waves
($\ell=2,4,\dots$) are dropped in this single-channel reduction.

For the S-wave component (O(4) isotropic amplitude $\phi(p)$ depending only
on $p \equiv |p_E|$), we define the dimensionless coupling:
\begin{equation}
  \alpha \equiv \frac{g^2}{16\pi^2},
  \label{eq:alpha_def}
\end{equation}
and integrate over the sphere $S^3$.
We use $\alpha$ only where the analytic Wick-Cutkosky spectrum
[Eq.~\eqref{eq:wc_exact}] is naturally expressed in it; the discretised
eigenvalue problem and all numerical tables use $g^2$, with the one-to-one
dictionary $\alpha=g^2/(16\pi^2)$ and $\lambda_{\max}=1/g^2_{\min}$, where
$\lambda_{\max}$ is the largest \emph{eigenvalue} of the discretised operator
(Sec.~\ref{sec:discretization}), \emph{not} a coupling.
The O(4) solid angle $\Omega_4$ divided by the momentum-space
measure factor $(2\pi)^4$ gives the prefactor
$\Omega_4/(2\pi)^4 = 1/(8\pi^2)$.
Writing $q \equiv |k_E|$ for the
radial magnitude of the Euclidean integration momentum, the resulting O(4)
S-wave BSE is:
\begin{equation}
  \DS(p,M)\,\phi(p)
  = \frac{g^2}{8\pi^2}\int_0^\infty dq\, q^3\,\KO(p,q;\mu)\,\phi(q),
  \label{eq:bse_1d}
\end{equation}
where the O(4)-projected inverse propagator product is obtained by taking the
O(4) angular average $\langle\cdot\rangle_{S^3}$ of
$(k_{1E}^2+m^2)(k_{2E}^2+m^2)$. Since
$k_{1E}^2 = p^2 + M^2/4 + Mp\cos\theta$ and
$k_{2E}^2 = p^2 + M^2/4 - Mp\cos\theta$, one finds:
\begin{equation}
  \begin{aligned}
    \left\langle (k_{1E}^2+m^2)(k_{2E}^2+m^2) \right\rangle_{S^3}
    &= \left(p^2+m^2+\frac{M^2}{4}\right)^{\!2} \\
    &\quad - M^2p^2\langle\cos^2\theta\rangle_{S^3},
  \end{aligned}
  \label{eq:D_avg}
\end{equation}
with $\langle\cos^2\theta\rangle_{S^3} = 1/4$ (exact on $S^3$ with uniform measure),
giving:
\begin{equation}
  \DS(p,M) = \left(p^2 + m^2 + \frac{M^2}{4}\right)^{\!2} - \frac{M^2 p^2}{4}.
  \label{eq:D_propagator}
\end{equation}
Note that $\DS(p,M) > 0$ for all $p \geq 0$ and all $M \geq 0$,
which can be seen by factoring
$\DS = \bigl[(p-M/2)^2+m^2\bigr]\bigl[(p+M/2)^2+m^2\bigr] > 0$;
the physical restriction $M < 2m$ for a bound state enters elsewhere
(threshold condition), not from the sign of $\DS$.
The O(4) S-wave exchange kernel, derived by angular integration over $S^3$
(see Appendix~\ref{app:kernel} and
Ref.~\cite{NieuwenhuisTjon1996}), is:
\begin{equation}
  \KO(p,q;\mu) = \frac{p^2+q^2+\mu^2 - \sqrt{[(p{-}q)^2{+}\mu^2][(p{+}q)^2{+}\mu^2]}}{2p^2q^2}.
  \label{eq:kernel_O4}
\end{equation}
In the massless limit $\mu \to 0$, Eq.~\eqref{eq:kernel_O4} reduces to the
Wick-Cutkosky kernel~\cite{Wick1954,Cutkosky1954}:
\begin{equation}
  \KO(p,q;0) = \frac{1}{\max(p^2,q^2)},
  \label{eq:wc_kernel}
\end{equation}
for which semi-analytic solutions serve as benchmarks~\cite{SchwartzZemach1966,Nakanishi1969,NieuwenhuisTjon1996}.

\subsection{\texorpdfstring{Wick-Cutkosky benchmark ($\mu = 0$)}{Wick-Cutkosky benchmark (mu = 0)}}

In the massless exchange limit $\mu \to 0$, the exchange kernel reduces to
$\KO(p,q;0) = 1/\max(p^2,q^2)$~\cite{Wick1954,Cutkosky1954}
(cf.\ Eq.~\eqref{eq:wc_kernel}). The \emph{full} angle-resolved equation
(without the O(4) angular-average approximation used in Sec.~\ref{sec:bse})
admits the exact eigenvalue
spectrum~\cite{Wick1954,Cutkosky1954,Nakanishi1969}
\begin{equation}
  \alpha_n^{\rm WC} = \frac{n(n+1)}{1 - \eta^2}, \qquad
  \eta = \frac{M}{2m}, \quad n = 1,2,3,\ldots
  \label{eq:wc_exact}
\end{equation}
exhibiting the O(4) degeneracy pattern characteristic of the Coulomb-like
interaction in four-dimensional Euclidean space. Our angular-averaged
single-channel equation deviates from this exact spectrum at finite $\eta$
(the deviation grows with $\eta$), so Table~\ref{tab:wc_benchmark}
contains numerical results from the present discretized, angular-averaged
formulation — not values from Eq.~\eqref{eq:wc_exact}.
A quantitative comparison with the exact formula is reserved for the
$\eta\to 0$ limit and the spectral ratio test below.

Table~\ref{tab:wc_benchmark} shows the minimum coupling $\alpha_{\min} =
g^2_{\min}/(16\pi^2)$ obtained numerically for $\mu\to 0$ and $\mu/m=0.15$
at several trial masses using the present angular-averaged formulation ($N=64$,
$p_{\max}=25\,m$).
The ratio $\alpha_{\min}(\mu=0.15)/\alpha_{\min}(\mu\to0)$ ranges from $1.038$
to $1.056$ across the tabulated masses, i.e., the exchange mass increases the
required coupling by $3.8\text{--}5.6\%$, consistent with the intuition that a
massive exchange boson provides less binding than a massless one.

\begin{figure}[tbp]
  \centering
  \includegraphics[width=1.0\textwidth]{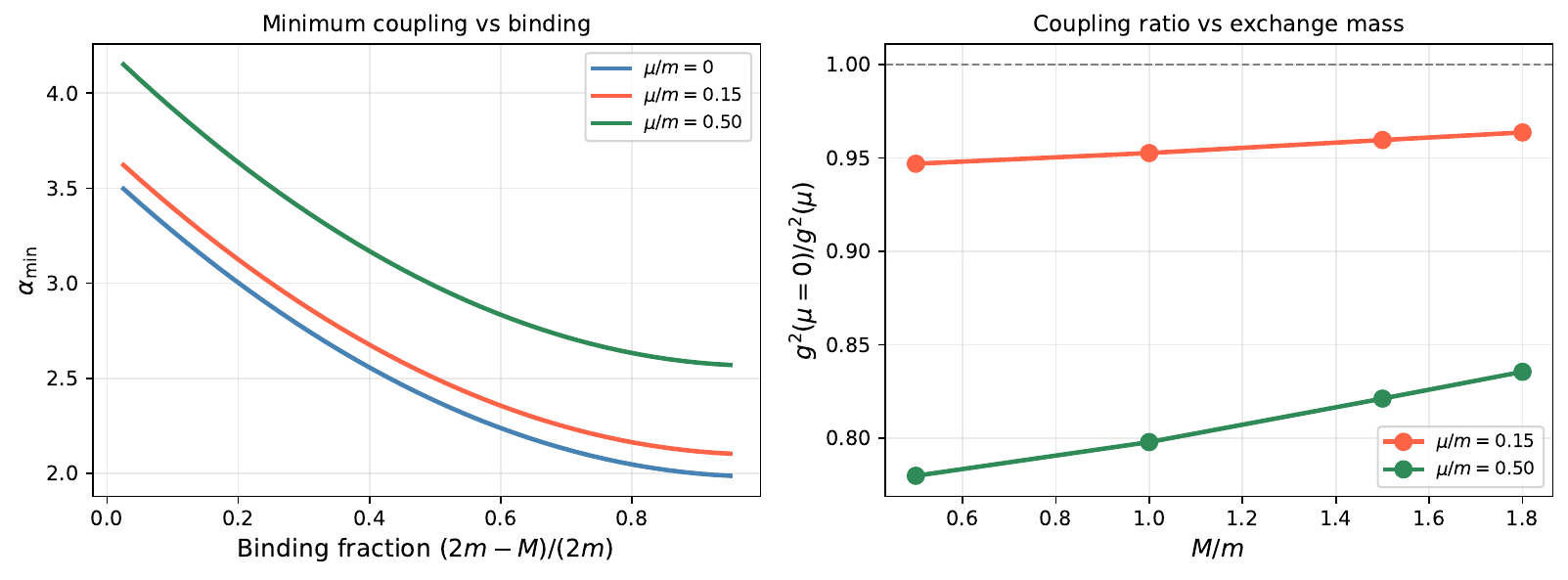}
  \caption{Wick-Cutkosky comparison ($N=64$, $p_{\max}=25\,m$). (Left) Minimum coupling $\alpha_{\min}$
    vs binding fraction $(2m-M)/(2m)$ for $\mu/m = 0, 0.15, 0.50$.
    More deeply bound states require \emph{less} coupling — a hallmark of
    the 4D relativistic BSE in the ladder approximation.
    (Right) Coupling ratio $g^2(\mu{=}0)/g^2(\mu) < 1$ plotted as a function of
    the binding fraction: values below unity confirm that a massive exchange boson
    requires $3.8\text{--}5.6\%$ \emph{more} coupling than the massless case
    (i.e.\ the inverse ratio $\alpha_{\min}(\mu)/\alpha_{\min}(\mu{=}0)
    \approx 1.038\text{--}1.056 > 1$, as listed in Table~\ref{tab:wc_benchmark}).}
  \label{fig:wc_benchmark}
\end{figure}

Figure~\ref{fig:wc_benchmark} illustrates this comparison for several values of the exchange mass $\mu$.

\begin{table}[tbp]
  \caption{Minimum coupling $\alpha_{\min}=g^2_{\min}/(16\pi^2)$ from
    the angular-averaged discretized BSE at mass $M/m$, for $\mu\to 0$ and
    $\mu/m=0.15$ ($N=64$ quadrature points, $p_{\max}=25\,m$). These are numerical results from
    the present formulation; they do not coincide with the exact Wick-Cutkosky
    spectrum Eq.~\eqref{eq:wc_exact}, which holds for the full angle-resolved
    equation.}
  \label{tab:wc_benchmark}
  \begin{tabular}{ccccc}
    \toprule
    $M/m$ & $\alpha_{\min}(\mu{=}0)$ & $\alpha_{\min}(\mu{=}0.15)$ &
    Ratio & $E_{\rm bind}/m$ \\
    \midrule
    0.50 & 2.083 & 2.200 & 1.056 & 1.50 \\
    1.00 & 2.382 & 2.500 & 1.050 & 1.00 \\
    1.50 & 2.879 & 3.001 & 1.042 & 0.50 \\
    1.80 & 3.273 & 3.396 & 1.038 & 0.20 \\
    \bottomrule
  \end{tabular}
\end{table}

For the deep-bound limit $\eta \to 0$ ($M \ll 2m$), the exact result
is $\alpha_1^{\rm WC} = 2$, and our discretization at $N=128$, $M/m=0.001$,
$\mu\to 0$ yields $\alpha_1 = 1.9949$ (relative error $0.26\%$), in
good agreement with the analytic Wick-Cutkosky value.
The remaining discrepancy reflects the finite IR cutoff $M/m=0.001>0$ and
the finite number of quadrature points.
At finite $\eta$, the full propagator product
$(p^2{+}m^2{+}M^2\!/4)^2 - M^2 p^2\cos^2\theta$ depends on
$\theta=\angle(\hat{p}_E,\hat{P}_E)$, and our code replaces
$\cos^2\theta$ by its O(4) average $1/4$. In the exact
Wick-Cutkosky solution the angular-dependent propagator couples the
$l=0,2,4,\ldots$ partial waves; our single-channel ($l=0$) equation
with the angular-averaged propagator introduces deviations
that grow toward threshold.

The eigenvalue \emph{ratios} converge to the exact O(4) pattern:
\begin{equation}
  \frac{\alpha_n}{\alpha_1} \;\to\; \frac{n(n+1)}{2},
  \label{eq:wc_ratios}
\end{equation}
with deviations $<2\%$ for $n \le 4$ at $N=128$. This confirms that
our discretization correctly reproduces the spectral \emph{structure} of the
Wick-Cutkosky model.

\subsection{Eigenvalue structure}

Equation~\eqref{eq:bse_1d} is a homogeneous Fredholm equation of the second kind.
A non-trivial solution $\phi(p)$ exists only for specific values of the pair
$(g^2, M)$: given a trial bound-state mass $M$, the minimum coupling $g^2_{\min}(M)$
satisfies the spectral condition of the integral operator.
In more general ladder-BSE settings, especially for excited states, the
interpretation of the spectrum can be complicated by abnormal states and even
complex eigenvalues~\cite{AhligAlkofer1999}. The present Euclidean O(4)
ground-state reduction avoids that complication by leading, after
discretization and symmetrization, to the real symmetric eigenvalue problem of
Sec.~\ref{sec:discretization}.

\section{\label{sec:discretization}Discretization and Symmetrization}

\subsection{Gauss-Legendre quadrature}

The radial integral in Eq.~\eqref{eq:bse_1d} extends over the semi-infinite
interval $q \in [0,\infty)$. For the numerical solution we truncate this to
$[0,p_{\max}]$ with a UV cutoff $p_{\max} = 20m$, which is sufficient for the
parameter range considered here and was checked explicitly for stability of the
largest eigenvalue. We then map the standard Gauss-Legendre nodes and weights
$\{\xi_i,\tilde{w}_i\}$ from $[-1,1]$ to the physical momentum interval
$[0,p_{\max}]$ via
\begin{equation}
  p_i = \frac{p_{\max}}{2}(\xi_i + 1), \quad
  w_i  = \frac{p_{\max}}{2}\tilde{w}_i, \quad
  i = 1,\ldots,N,
\end{equation}
where $\{\xi_i, \tilde{w}_i\}$ are standard Gauss-Legendre nodes and weights
on $[-1,1]$~\cite{NumericalRecipes2007}. With this choice, the radial measure
$q^3 dq$ in Eq.~\eqref{eq:bse_1d} is represented by the discrete weights
$w_i p_i^3$, so that the integral equation is reduced directly to a finite-
dimensional matrix problem. This type of quadrature discretization is standard
in Euclidean Bethe-Salpeter studies; see, e.g.,
Refs.~\cite{SchwartzZemach1966,Bauhoff1974,NieuwenhuisTjon1996}.

\subsection{Generalized eigenvalue problem}

Evaluating Eq.~\eqref{eq:bse_1d} at the quadrature points $p_i$ and replacing
the integral by the Gauss-Legendre sum gives
\begin{equation}
  D_i\,\phi_i = \frac{g^2}{8\pi^2}\sum_{j=1}^N w_j\, p_j^3\,\KO(p_i,p_j;\mu)\,\phi_j,
  \label{eq:disc_sum}
\end{equation}
with
\begin{equation}
  D_i \equiv \DS(p_i,M).
  \label{eq:Di_def}
\end{equation}
It is convenient to introduce the $g^2$-independent kernel matrix
\begin{equation}
  \tilde{K}^{(0)}_{ij} = \frac{w_j\, p_j^3}{8\pi^2}\,\KO(p_i,p_j;\mu),
  \label{eq:K0disc}
\end{equation}
so that the discretized equation becomes
\begin{equation}
  D_i\,\phi_i = g^2 \sum_j \tilde{K}^{(0)}_{ij}\,\phi_j.
  \label{eq:gevp_component}
\end{equation}
In matrix form this is the generalized eigenvalue problem
\begin{equation}
  \tilde{\mathbf{K}}^{(0)}\,\boldsymbol{\phi} = \frac{1}{g^2}\,\mathbf{D}\,\boldsymbol{\phi},
  \qquad
  \mathbf{D}=\mathrm{diag}(D_i),
  \label{eq:gevp_matrix}
\end{equation}
with eigenvalue $1/g^2$.

\subsection{Symmetric standard eigenvalue problem}

The matrix $\tilde{\mathbf{K}}^{(0)}$ is not symmetric because the quadrature
weight appears only on the index $j$. To obtain a symmetric problem, we absorb
the radial measure into the wavefunction by defining
\begin{equation}
  \psi_i = \sqrt{w_i p_i^3}\,\phi_i.
  \label{eq:psi_def}
\end{equation}
Then Eq.~\eqref{eq:gevp_matrix} becomes
\begin{equation}
  [K_{\rm sym}]_{ij} = \frac{\sqrt{w_i p_i^3\, w_j p_j^3}}{8\pi^2}\,\KO(p_i,p_j;\mu).
  \label{eq:Ksym}
\end{equation}
which is manifestly real and symmetric. The GEVP therefore reads
\begin{equation}
  \mathbf{K}_{\rm sym}\,\boldsymbol{\psi} = \frac{1}{g^2}\,\mathbf{D}\,\boldsymbol{\psi}.
  \label{eq:gevp_sym}
\end{equation}
Finally, because $\mathbf{D}$ is diagonal and positive for $M<2m$, we may apply
the congruence transform
\begin{equation}
  \boldsymbol{v} = \mathbf{D}^{1/2}\boldsymbol{\psi}.
  \label{eq:chi_def}
\end{equation}
Multiplying Eq.~\eqref{eq:gevp_sym} from the left by $\mathbf{D}^{-1/2}$ gives
\begin{equation}
  H_{\rm sym}\,\bm{v} = \frac{1}{g^2}\,\bm{v}, \quad
  H_{\rm sym} = \bm{D}^{-1/2} K_{\rm sym} \bm{D}^{-1/2}.
  \label{eq:Hsym_eigenproblem}
\end{equation}
$H_{\rm sym}$ is real and symmetric. Its
\emph{maximum eigenvalue} $\lambda_{\max}(M)$ determines the minimum
coupling: $g^2_{\min}(M) = 1/\lambda_{\max}(M)$.

Choosing $N = 2^n$ identifies the $N$-dimensional coefficient
space with the Hilbert space of $n$ qubits, so the eigenvector $\bm{v}$ is stored
in the amplitudes of an $n$-qubit register. Because $H_{\rm sym}$ is real
symmetric (hence Hermitian), it can be represented as an operator on this space
and expanded in the $n$-qubit Pauli basis (Sec.~\ref{sec:quantum}), placing its
extremal eigenpair within reach of quantum eigensolvers such as the VQE.

\section{\label{sec:quantum}Quantum Computing Approach}

The strategy of this section is variational, in the same spirit as the
Rayleigh--Ritz method familiar from quantum field theory: rather than
diagonalizing the exponentially large matrix $H_{\rm sym}$ directly, we store its
eigenvector in the amplitudes of an $n$-qubit register, generate a family of
trial states $|\psi(\bm{\theta})\rangle$ with a shallow parametrized circuit, and
minimize the expectation value $\langle\psi(\bm{\theta})|\hat H_{\vqe}|
\psi(\bm{\theta})\rangle$ over the circuit angles $\bm{\theta}$. The following
three subsections treat, in turn, how the operator is cast into a
hardware-measurable form, how the trial state is constructed, and how the
minimization is performed. Readers with a field-theory background may find it
useful to picture the $n$-qubit register as a chain of $n$ spin-$\tfrac12$ sites
whose $2^{n}$-dimensional Hilbert space stores the discretized momentum-space
wave function of Sec.~\ref{sec:discretization}.

\subsection{Pauli decomposition}

Any $2^n \times 2^n$ Hermitian matrix can be expanded in the $n$-qubit
Pauli basis~\cite{Nielsen2000}:
\begin{equation}
  \hat{H} = \sum_{k=1}^{4^n} c_k\, \hat{P}_k, \quad
  c_k = \frac{1}{2^n}\,\mathrm{Tr}\!\left[\hat{P}_k\,\hat{H}\right],
  \label{eq:pauli_decomp}
\end{equation}
where $\hat{P}_k \in \{I,X,Y,Z\}^{\otimes n}$. The $4^{n}$ Pauli
strings form an orthogonal basis for the $2^{n}\times 2^{n}$ Hermitian matrices
under the Hilbert--Schmidt product $\mathrm{Tr}[\hat P_j\hat P_k]=2^{n}
\delta_{jk}$, so Eq.~\eqref{eq:pauli_decomp} is an exact rewriting of the
operator---the analogue of expanding a field in a complete set of modes. Its
practical value is that each Pauli string is a directly measurable observable on
qubit hardware, so the energy is obtained as the weighted sum $\langle\hat
H\rangle=\sum_k c_k\langle\hat P_k\rangle$, and the number of nonzero terms sets
the measurement cost of the algorithm. Since $H_{\rm sym}$ is real
symmetric, all coefficients $c_k$ with odd numbers of $Y$ operators vanish,
because such Pauli strings are purely imaginary. By contrast, strings built
from $I$, $X$, and $Z$ do contribute, and Pauli strings containing an
\emph{even} number of $Y$ operators can also have nonzero coefficients.
Thus the real-symmetric structure does not single out $Y$ operators alone; it
eliminates only the odd-$Y$ sector, leaving a Hamiltonian composed of
$I/X/Z$ strings together with even-$Y$ combinations. For the $n=4$ example
this yields exactly $(4^4+2^4)/2 = 136$ nonzero Pauli terms out of $4^4=256$,
a fraction of $0.531$; the general formula $(4^n+2^n)/2$ is derived and
verified numerically.

We define $\hat{H}_{\vqe} = -H_{\rm sym}$ so that minimizing the VQE
expectation value finds $-\lambda_{\max}(H_{\rm sym})$, i.e.,
\begin{equation}
  \min_{\bm{\theta}}\langle\psi(\bm{\theta})|\hat{H}_{\vqe}|\psi(\bm{\theta})\rangle
  = -\lambda_{\max}(H_{\rm sym}) = -1/g^2_{\min}.
  \label{eq:vqe_min}
\end{equation}
The sign flip turns the largest eigenvalue of $H_{\rm sym}$ into the lowest
(most negative) eigenvalue of $\hat H_{\vqe}$, so Eq.~\eqref{eq:vqe_min} is
nothing but the standard variational principle: for any normalized trial state
the expectation value bounds the lowest eigenvalue of $\hat H_{\vqe}$ from above,
and the bound is saturated when $|\psi(\bm{\theta})\rangle$ reaches the target
eigenvector. Minimizing over the circuit angles $\bm{\theta}$ therefore returns
$-\lambda_{\max}(H_{\rm sym})$ and, through $g^2_{\min}=1/\lambda_{\max}$, the
minimum coupling that supports a bound state at mass $M$.

\subsection{MPS tensor-network ansatz}

The variational state is built to approximate the dominant eigenvector of
$H_{\rm sym}$ in the computational basis. After normalizing the discretized
amplitude $\bm{v}$ from
Eq.~\eqref{eq:Hsym_eigenproblem}, we encode it as
an $n$-qubit state
\begin{equation}
  |v\rangle = \sum_{i=0}^{N-1} v_i |i\rangle,
  \qquad N=2^n,
  \label{eq:state_encoding}
\end{equation}
where the basis state $|i\rangle$ labels the $i$th quadrature point.Concretely, $|i\rangle$ is the computational-basis state whose binary label is
the integer $i$, i.e.\ a joint eigenstate of the single-qubit $Z$ operators;
the register thus holds all $N=2^{n}$ discretized amplitudes simultaneously, an
exponentially compact encoding in which $n$ qubits carry $2^{n}$ real numbers.
The
task of the ansatz is therefore to generate a real-amplitude state whose
coefficients reproduce the smooth momentum-space profile of the BSE ground
state.

The physical motivation for using an MPS-type circuit is the \emph{area law}
for entanglement entropy~\cite{White1992,Vidal2003,Schollwoeck2011,Orus2014}.
After the O(4) projection the amplitude depends on a single radial variable,
and the corresponding qubit state shows a rapidly decaying Schmidt spectrum
across bipartitions. The Schmidt spectrum across a cut is the set
of eigenvalues of the reduced density matrix obtained by tracing out the
complementary qubits, and its von Neumann entropy $S=-\sum_a\lambda_a\ln\lambda_a$
is the entanglement entropy across that cut---the same quantity studied for
ground states in lattice field theory and conformal field theory. An area law
means this entropy stays bounded as the register grows, which is exactly the
regime in which a small bond dimension suffices. In other words, the target state is not expected to
require strong long-range entanglement across the whole register. This is
verified explicitly in Sec.~\ref{sec:results}, where the von Neumann entropy
remains far below its maximal value and bond dimension $\chi=2$ already gives
very high fidelity.

The ansatz in the following is a shallow
nearest-neighbor circuit evaluated on PennyLane's \texttt{default.tensor}
backend in MPS mode \cite{PennyLane2024}. 
One first applies a Hadamard gate to every qubit,
producing a delocalized initial state with support on all computational-basis
configurations. Explicitly, $H^{\otimes n}|0\cdots0\rangle=
N^{-1/2}\sum_{i=0}^{N-1}|i\rangle$ is the uniform superposition---a flat,
featureless momentum profile that the subsequent entangling blocks reshape into
the peaked BSE ground state. One then performs $L$ sequential sweeps of local two-qubit
updates along the 1D qubit chain. A \emph{sweep} applies the two-qubit
block $U_{\ell,b}$ to every nearest-neighbor bond $b=0,1,\dots,n-2$ in turn,
from one end of the chain to the other; $L$ such sweeps are then stacked. Here
$R_Y(\theta)=\exp(-\mathrm{i}\theta Y/2)$ is a single-qubit rotation about the
$Y$ axis, and $\mathrm{CNOT}_{b,b+1}$ is the controlled-NOT gate with control
$b$ and target $b+1$. For sweep $\ell$ and bond $b=0,\dots,n-2$,
the local block is
\begin{equation}
  U_{\ell,b} = \mathrm{CNOT}_{b,b+1}
  \bigl[R_Y(\theta^{(1)}_{\ell,b})_b \otimes
        R_Y(\theta^{(2)}_{\ell,b})_{b+1}\bigr].
\end{equation}
The full trial state is therefore
\begin{equation}
  |\psi(\bm{\theta})\rangle =
  \prod_{\ell=1}^{L} \prod_{b=0}^{n-2} U_{\ell,b}
  H^{\otimes n}|0\cdots 0\rangle.
  \label{eq:mps_ansatz_state}
\end{equation}
This construction mirrors the structure of a matrix-product state: each block
introduces only local entanglement between neighboring qubits, while repeated
sweeps propagate correlations across the full register. Concretely,
a matrix-product state writes the $2^{n}$ amplitudes $v_{s_1\cdots s_n}$ as a
chain of matrix products $A^{s_1}\!\cdots A^{s_n}$, one factor per qubit, so the
full amplitude tensor is described by only $O(n\chi^{2})$ numbers rather than
$2^{n}$, with $\chi$ (defined below) setting how much correlation is retained.
This is the same variational class that underlies the density-matrix
renormalization group.

The exclusive use of $R_Y$ rotations is also intentional. Because $H_{\rm sym}$
is real and symmetric, its dominant eigenvector may be chosen real in the
computational basis. The target state therefore does not require arbitrary
complex phases. Starting from the real state $H^{\otimes n}|0\cdots0\rangle$,
the gate set \{$R_Y$, CNOT\} generates a broad family of real-amplitude trial
states, which is precisely the sector relevant for the present BSE problem.
Including additional $R_X$ or $R_Z$ rotations would enlarge the ansatz to
complex-valued states, but that extra freedom is not needed here and would add
variational parameters without a clear physical motivation.

For a register of $n$ qubits and $L$ sweeps, the number of variational
parameters is
\begin{equation}
  N_{\rm par} = 2(n-1)L.
\end{equation}
Here $L$ is the circuit depth in units of nearest-neighbor sweeps; it is not
the MPS bond dimension $\chi$. The bond dimension
$\chi$ of a matrix-product state is the size of the auxiliary (virtual) index
that links neighboring sites; equivalently, across any bipartition of the chain
an MPS of bond dimension $\chi$ can represent exactly those states whose Schmidt
rank does not exceed $\chi$. Since the Schmidt rank is the minimum number of
product terms needed to write the state across that cut, $\chi$ is precisely the
minimal bond dimension for a truncation-free representation, and a state with a
rapidly decaying Schmidt spectrum is already well approximated by small $\chi$
(quantified in Sec.~\ref{sec:results}). Increasing $L$ enlarges the variational family
by adding more local update layers, whereas $\chi$ measures the amount of
bipartite entanglement that an MPS representation can carry. The two are
related only indirectly: larger $L$ can produce states that may require a
larger effective bond dimension for an exact MPS description, but in general
$L \neq \chi$.
The implementation uses $L=2$, so for the main $N=16$ example
($n=4$ qubits) the circuit contains $12$ variational parameters. This choice
is motivated by the observed low-entanglement structure of the target state, so
we start from a deliberately modest circuit rather than a deep
hardware-efficient ansatz. However, low entanglement alone does not prove that
$L=2$ is already optimal: establishing that would require an explicit
layer-depth convergence study showing that larger $L$ values do not improve the
best variational energy within the target accuracy. We therefore interpret
$L=2$ as a pragmatic minimal choice used in the present implementation, not as
a universal or rigorously optimized depth. A dedicated depth scan with the same
ansatz family and ADAM settings for $L=1,\dots,8$ (see below in Fig.~\ref{fig:vqe_depth_scan})
shows that the dominant
improvement already occurs between $L=1$ and $L=2$ ($\Delta\lambda_{\max}
\approx 9.7\times 10^{-5}$ for the $N=16$ example), while further increases in
$L$ change $\lambda_{\max}$ only at the $10^{-7}$ level; this supports the use
of a shallow circuit for the present low-entanglement problem.
On the tensor-network backend we allow a maximum internal bond dimension
$\chi_{\rm max}=16$, which is well above the exact requirement for $n=4$ and
therefore does not constrain the state representation. The role of the MPS
ansatz is thus not to enforce an artificial truncation for the small systems
studied here, but to provide a physically motivated circuit architecture that
remains scalable when $N$ is increased.

\subsection{VQE algorithm}

The VQE algorithm minimizes the expectation value of $\hat H_{\vqe}=-H_{\rm sym}$.
Within the PennyLane framework~\cite{PennyLane2024} the tensor-network contraction
is carried out with the \texttt{quimb}
library~\cite{quimb2018}. For this, the implementation uses the
MPS mode with maximum bond dimension $\chi_{\max}=16$, which is comfortably
above the exact requirement for the $n=4$ demonstration problem, as shown below.

Gradients are evaluated with the parameter-shift rule~\cite{Mitarai2018},
which returns the exact analytic derivative of the expectation value with respect
to each $R_Y$ angle by evaluating the same circuit at two shifted parameter
values, $\partial_\theta\langle\hat H\rangle=\tfrac12[\langle\hat
H\rangle_{\theta+\pi/2}-\langle\hat H\rangle_{\theta-\pi/2}]$, thereby avoiding
finite-difference error, and
the circuit parameters are initialized randomly in the interval
$[-\pi,\pi]$ with a fixed seed for reproducibility. 
The optimizer is ADAM~\cite{Kingma2015}
with step size $0.03$. The optimization loop allows at most $1000$ update
steps, but in practice terminates earlier once the change in the objective
between successive iterations satisfies
\begin{equation}
  |E^{(t)} - E^{(t-1)}| < 10^{-8},
\end{equation}
after an initial burn-in of $80$ iterations.

For the production results reported in this paper, we optimize the same
variational energy functional
\begin{equation}
  E(\bm{\theta}) = \langle \psi(\bm{\theta}) | \hat H_{\vqe} | \psi(\bm{\theta}) \rangle
\end{equation}
with a classical L-BFGS-B routine, evaluating the expectation value exactly on
the corresponding statevector/tensor-network representation. This is therefore
not a different variational principle, but the noiseless Rayleigh-Ritz problem
associated with the chosen ansatz, solved with a classical optimizer rather
than through repeated quantum-circuit evaluations. The explicit PennyLane VQE
implementation was used as a validation of this objective and reproduces the
same optimum to numerical precision for the benchmark cases considered here
when executed with PennyLane $\geq 0.37$~\cite{PennyLane2018,Cerezo2021,Kandala2017}.
Accordingly, the Pauli-measurement overhead discussed later in
Sec.~\ref{sec:outlook} is not a cost of the present tensor-network simulation;
it is a prospective cost for realizing the same variational objective on
sampling-based gate hardware.

\section{\label{sec:results}Results}

\subsection{Classical diagonalization}

Table~\ref{tab:results} shows $\lambda_{\max}(M)$ and $g^2_{\min}(M)$
for $\mu/m = 0.15$ at several trial masses.
As $M\to2m$ the binding energy vanishes and the required coupling increases
(Table~\ref{tab:results}). In the \emph{exact}, fully angle-resolved
Wick-Cutkosky spectrum the ground-state coupling \emph{diverges} at threshold,
$\alpha_1=2/(1-\eta^2)\to\infty$ as $\eta=M/2m\to1$ [Eq.~\eqref{eq:wc_exact}];
it is thus the coupling that diverges here -- the relativistic counterpart of,
and opposite in direction to, the non-relativistic threshold divergence of the
scattering length. The present single-channel O(4) S-wave average does
\emph{not} reproduce this divergence: it omits the $\ell=2,4,\dots$ partial
waves that drive it, so the computed $g^2_{\min}$ instead \emph{saturates} to a
finite value near threshold for both $\mu\to0$ and $\mu/m=0.15$.
Figure~\ref{fig:amplitude} (left) shows this explicitly by overlaying the exact
spectrum; the gap that opens toward threshold is the same single-channel
deviation noted in Sec.~\ref{sec:bse}. This also explains why the plotted curve
approaches a finite value rather than diverging.

\begin{table}[tbp]
  \caption{Classical BSE results: maximum eigenvalue $\lambda_{\max}$,
    minimum coupling $g^2_{\min} = 1/\lambda_{\max}$, and binding energy
    $E_{\rm bind}/m = 2-M/m$, for $\mu/m=0.15$, $N=16$.}
  \label{tab:results}
  \begin{tabular}{cccc}
    \toprule
    $M/m$ & $\lambda_{\max}$ & $g^2_{\min}$ & $E_{\rm bind}/m$ \\
    \midrule
    1.2 & $2.4325\times10^{-3}$ & 411.09 & 0.80 \\
    1.4 & $2.2563\times10^{-3}$ & 443.21 & 0.60 \\
    1.6 & $2.0835\times10^{-3}$ & 479.97 & 0.40 \\
    1.8 & $1.9178\times10^{-3}$ & 521.44 & 0.20 \\
    \bottomrule
  \end{tabular}
\end{table}

\subsection{Bound state amplitude}

\begin{figure}[tbp]
  \centering
  \includegraphics[width=1.0\textwidth]{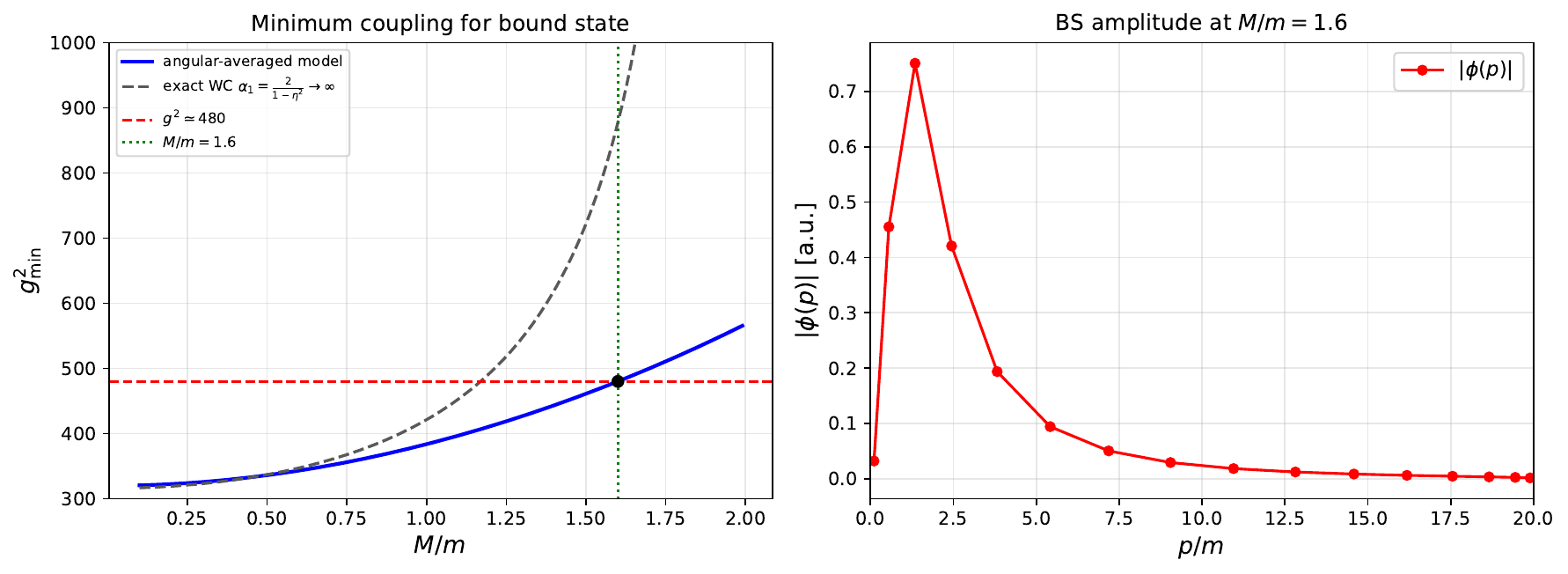}
  \caption{(Left) Minimum coupling $g^2_{\min}(M)$ versus the
    trial bound-state mass $M/m$, for $\mu/m=0.15$ (solid). The horizontal
    dashed line shows the reference coupling $g^2\simeq 480$ corresponding
    to the benchmark point $M/m=1.6$; its intersection with the curve marks
    that mass. The single-channel result \emph{saturates} toward
    threshold, whereas the exact angle-resolved Wick-Cutkosky coupling
    $\alpha_1=2/(1-\eta^2)$ (gray dashed) \emph{diverges} as $M\to2m$. (Right) Euclidean BS
    amplitude $|\phi(p)|$ at $M/m=1.6$.}
  \label{fig:amplitude}
\end{figure}

Figure~\ref{fig:amplitude} summarizes the classical solution at $M/m=1.6$.
Its left panel shows the minimum-coupling curve $g^2_{\min}(M)$ together with
the horizontal reference line at $g^2\simeq 480$, chosen to match the
demonstration point $M/m=1.6$. Its intersection with $g^2_{\min}(M)$ therefore
marks the same bound-state mass displayed in the right panel. The right panel shows the Euclidean BS
amplitude $|\phi(p)|$ at $M/m=1.6$. On the discretized momentum grid used here,
the amplitude rises from near zero at small $p$, reaches a maximum near
$p\approx 1.3\,m$, and then decreases monotonically at large momenta.

\subsection{VQE accuracy}

For $n=4$ qubits ($N=16$), the VQE achieves a mean relative error
\begin{equation}
  \overline{\delta}_{\rm rel} =
  \frac{|\lambda_{\max}^{\vqe} - \lambda_{\max}^{\rm cl}|}
       {\lambda_{\max}^{\rm cl}} < 1\%,
\end{equation}
with the best grid point $(L,\eta)=(2,0.01)$ yielding
$\overline{\delta}_{\rm rel} = 4.8\times 10^{-3}$
(Table~\ref{tab:vqe_grid_summary}),
after convergence in $\lesssim 300$ ADAM steps.
The MPS bond dimension $\chi = 2$ already reproduces $> 99\%$ of the
ground-state fidelity, confirming area-law entanglement.

\begin{figure}[ht]
  \includegraphics[width=1.0\textwidth]{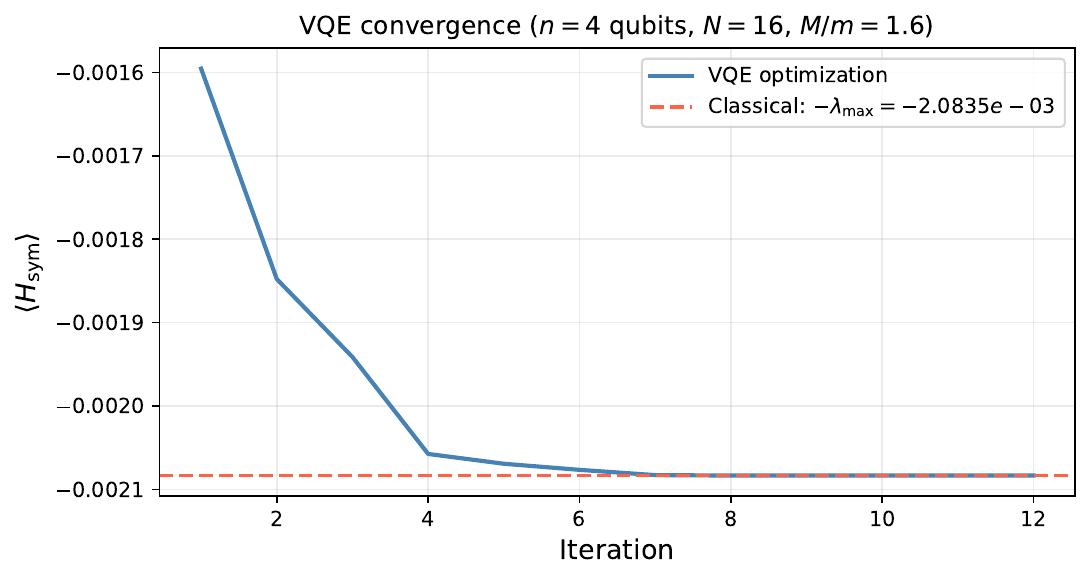}
  \caption{Rayleigh-quotient optimization (L-BFGS-B) on $H_{\rm sym}$
    for $n=4$ qubits ($N=16$), $M/m = 1.6$, $\mu/m = 0.15$,
    illustrating the cost-function landscape.
    The horizontal dashed line is the classical eigenvalue
    $-\lambda_{\max} = -2.0835\times10^{-3}$.
    The ADAM-based VQE (Table~\ref{tab:vqe_grid_summary}) converges
    in $\lesssim 300$ steps to the same target.}
  \label{fig:convergence}
\end{figure}

To assess optimization robustness, we performed a small sweep over ADAM step
sizes $\eta \in \{0.01,0.03\}$ and ansatz depths $L \in \{1,2\}$, using five
random initializations per grid point. Table~\ref{tab:vqe_grid_summary}
shows that increasing the depth from $L=1$ to $L=2$ lowers the mean relative
error from approximately $5\times 10^{-2}$ to below $10^{-2}$. Within the
tested set, the most accurate and stable point is $(L,\eta)=(2,0.01)$, with
mean relative error $4.803\times 10^{-3}$ and the smallest observed spread in
$\lambda_{\max}^{\vqe}$.

\begin{table}[tbp]
  \caption{Hyperparameter sweep for the $N=16$ ($n=4$ qubit) VQE benchmark.
    Each entry reports the mean and sample standard deviation over five random
    initializations for the ansatz depth $L$ and ADAM step size $\eta$.}
  \label{tab:vqe_grid_summary}
  \begin{tabular}{cccccc}
    \toprule
    $L$ & $\eta$ & $\overline{\lambda}_{\max}^{\vqe}$ & $\sigma(\lambda_{\max}^{\vqe})$ &
    $\overline{\delta}_{\rm rel}$ & $\sigma(\delta_{\rm rel})$ \\
    \midrule
    1 & 0.010 & $1.977\times10^{-3}$ & $2.002\times10^{-5}$ & $5.120\times10^{-2}$ & $9.608\times10^{-3}$ \\
    1 & 0.030 & $1.976\times10^{-3}$ & $2.046\times10^{-5}$ & $5.157\times10^{-2}$ & $9.820\times10^{-3}$ \\
    2 & 0.010 & $2.073\times10^{-3}$ & $1.219\times10^{-5}$ & $4.803\times10^{-3}$ & $5.851\times10^{-3}$ \\
    2 & 0.030 & $2.068\times10^{-3}$ & $2.089\times10^{-5}$ & $7.452\times10^{-3}$ & $1.002\times10^{-2}$ \\
    \bottomrule
  \end{tabular}
\end{table}

\begin{figure}[tbp]
  \includegraphics[width=1.0\textwidth]{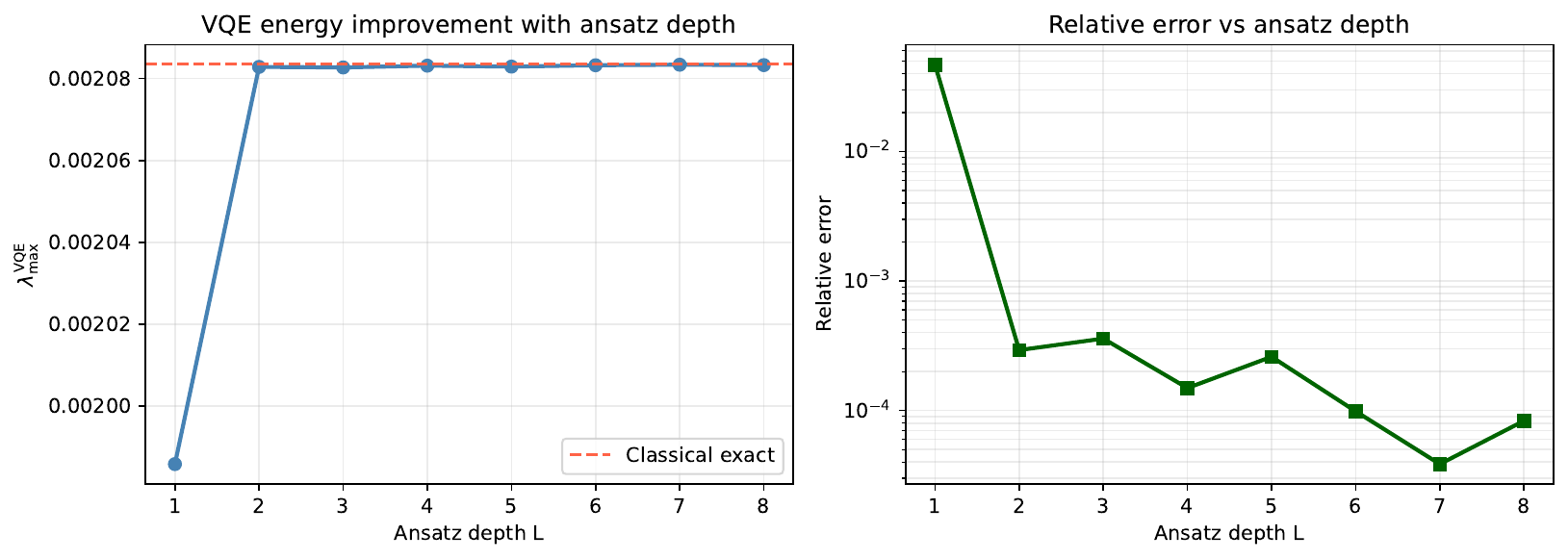}
  \caption{VQE ansatz depth scan for $N=16$ ($n=4$ qubits). (Left)
    $\lambda_{\max}^{\vqe}$ as a function of circuit depth $L$; the
    horizontal dashed line marks the classical exact value. (Right)
    Relative error on a logarithmic scale. The dominant improvement
    occurs from $L=1$ to $L=2$; deeper circuits yield only marginal gains.}
  \label{fig:vqe_depth_scan}
\end{figure}

To further investigate the role of circuit depth, we scan the ansatz depth
from $L=1$ to $L=8$ using the same gate family and ADAM optimizer, with
warm-starting from the previous depth's optimized parameters.
Figure~\ref{fig:vqe_depth_scan} shows the result: the dominant improvement
occurs between $L=1$ and $L=2$ ($\Delta\lambda_{\max}\approx
9.7\times10^{-5}$), while further increases change $\lambda_{\max}$ only at
the $10^{-7}$ level. This confirms that $L=2$ already saturates the accessible
variational minimum for this low-entanglement problem.

\subsection{Entanglement analysis}

To make the MPS argument quantitative, we analyze the entanglement structure
of the normalized classical ground state $|v\rangle$ obtained from
Eq.~\eqref{eq:Hsym_eigenproblem}. For a bipartition
$[1,\dots,k]\,|\,[k{+}1,\dots,n]$ of the $n=4$ qubit register, the statevector
is reshaped into a $2^k \times 2^{n-k}$ coefficient matrix $C^{(k)}$ and
subjected to a singular-value decomposition,
\begin{equation}
  C^{(k)} = U\,\mathrm{diag}\!\bigl(s_1^{(k)},s_2^{(k)},\dots\bigr)\,V^\dagger,
\end{equation}
which is the Schmidt decomposition of the state across that cut. The reduced
density matrix of the first block is then
\begin{equation}
  \rho_k = C^{(k)} C^{(k)\dagger},
  \qquad
  \mathrm{spec}(\rho_k)=\{(s_a^{(k)})^2\},
\end{equation}
so that the bipartite von Neumann entropy is
\begin{equation}
  S_{\rm vN}^{(k)} = -\sum_a (s_a^{(k)})^2 \log_2 (s_a^{(k)})^2.
\end{equation}
In an MPS language, the exact Schmidt rank across a cut is the bond dimension
required to represent the state without truncation, while the truncated
fidelity retained by bond dimension $\chi$ is
\begin{equation}
  F_{\chi}^{(k)} = \sum_{a=1}^{\chi} (s_a^{(k)})^2.
\end{equation}
These are the standard diagnostics connecting low entanglement to MPS
efficiency~\cite{White1992,Vidal2003,Schollwoeck2011,Orus2014,Eisert2010area}.

\begin{figure}[tbp]
  \includegraphics[width=1.0\textwidth]{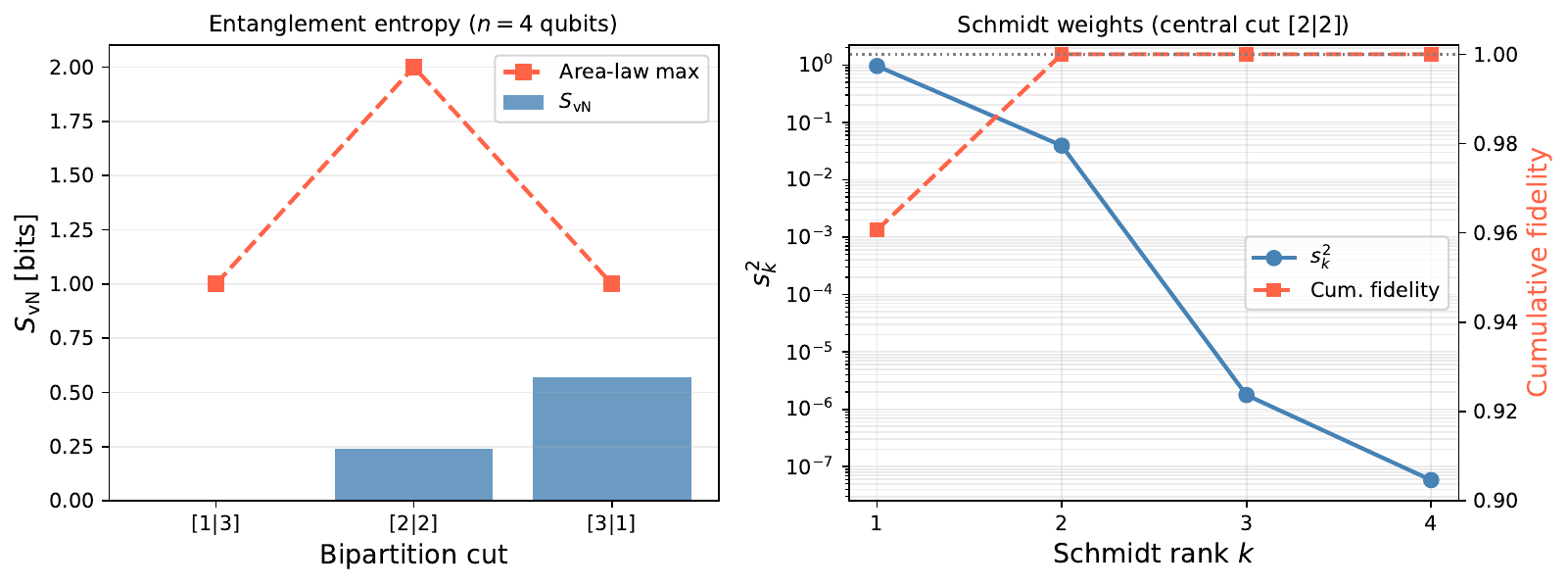}
  \caption{(Left) Von Neumann entanglement entropy $S_{\rm vN}$ for all
    bipartitions of the 4-qubit BSE ground state, far below the area-law
    maximum (dashed). (Right) Schmidt weights $s_k^2$ and cumulative
    fidelity at the central bipartition $[2|2]$: bond dimension $\chi=2$
    suffices for $99.9998\%$ fidelity, confirming the MPS ansatz is optimal.}
  \label{fig:entanglement}
\end{figure}

Figure~\ref{fig:entanglement} makes this logic explicit. Its left panel shows
the entanglement entropies for all three nontrivial bipartitions of the 4-qubit
state,
\begin{equation}
  S_{[1|3]} = 0.0047,\quad S_{[2|2]} = 0.2393,\quad S_{[3|1]} = 0.5675\;\text{bits},
\end{equation}
all far below the maximal value of $\min(k,n-k)$ bits, which is $1$ bit for the
$[1|3]$ and $[3|1]$ cuts and $2$ bits for the central $[2|2]$ cut. The right
panel of Fig.~\ref{fig:entanglement} then resolves the Schmidt spectrum of the
central bipartition $[2|2]$, which is the most demanding cut for an MPS
in the sense that it has the largest possible Schmidt rank ($\min(2^2,2^2)=4$);
the $[3|1]$ cut, despite having a higher entropy ($S_{[3|1]}=0.5675$~bits),
is geometrically limited to rank $\min(2^3,2^1)=2$, so bond dimension $\chi=2$
is trivially exact there.
There the first Schmidt value already dominates strongly, so bond dimension
$\chi=1$ retains $96.1\%$ of the norm at the $[2|2]$ cut, while $\chi=2$ raises the cumulative
fidelity to $99.9998\%$.

The conclusion is therefore not just qualitative but algorithmic: the BSE
ground state in this qubit encoding has a rapidly decaying Schmidt spectrum and
very small bipartite entropies, so only a very small bond dimension is needed
to capture the physically relevant state. This is precisely the structure
expected of a one-dimensional area-law state and is the central reason why the
MPS ansatz is so effective for the Euclidean O(4) S-wave problem studied here.

\subsection{Scaling analysis}

The cost of the present formulation is compared against classical methods in the
consolidated Fig.~\ref{fig:quantum_advantage} of Sec.~\ref{sec:outlook}; we use a
single, consistent cost model there. Dense diagonalization scales as
$\mathcal{O}(N^3)$, whereas for the area-law ground state found here a classical
low-bond-dimension MPS/DMRG treatment scales as $\mathcal{O}(\chi^3 N)$ (and a
Lanczos iteration as $\mathcal{O}(kN)$). Crucially, this does not include the
$\mathcal{O}(4^n)$ Pauli-measurement overhead required on actual gate hardware
(see Sec.~\ref{sec:outlook}). The favourable MPS scaling should not be
interpreted as a genuine quantum advantage
for the present Euclidean O(4) S-wave problem. The reason is that here the
ground state obeys an area law, so the required MPS bond dimension stays small
as $N$ increases and classical MPS/Lanczos methods remain efficient; in
addition, an actual VQE implementation must still pay the Pauli-measurement and
optimization costs discussed in Sec.~\ref{sec:outlook}. The better scaling is
therefore only a proxy estimate for situations in which the effective problem
becomes genuinely higher-dimensional, such as direct Minkowski-space BSEs
(requiring discretization of two independent momentum variables) or
coupled-channel systems, where the extra correlations can force $\chi$ to grow
and remove the classical low-entanglement advantage.

The D-Wave approach of Ref.~\cite{Fornetti2024annealer} formulated the BSE
as a QUBO problem and obtained results for matrix sizes up to $64\times64$.
The physical parameters (masses, coupling constant) transfer directly between
the two formulations. However, the absolute eigenvalues differ
significantly—by roughly a factor of $4$ at the strongly bound benchmark
point $M/m=1.0$—because the angular-averaged Euclidean propagator
(Sec.~\ref{sec:bse}) couples partial waves differently from the exact
Minkowski-space treatment used in Ref.~\cite{Fornetti2024annealer}.
At strong binding ($\eta=M/(2m)=0.5$) the O(4) average
$\langle\cos^2\theta\rangle=1/4$ is only a rough approximation to the
fully angle-resolved propagator, so large deviations are expected.
The remaining difference is also affected by the different
basis functions (Gauss--Legendre quadrature vs.\ Nakanishi weights) and
vanishes in the continuum limit.
Our gate-based VQE approach is complementary: it targets the same eigenvalue
but uses the variational principle on a parameterized quantum circuit, which
allows for systematic improvement via richer ans\"atze and is more readily
extended to fault-tolerant algorithms.

\section{\label{sec:largeN}\texorpdfstring{Larger Discretizations: $N=32$ and $N=64$}{Larger Discretizations: N=32 and N=64}}

\subsection{Physical convergence}

This subsection addresses a different question from Table~\ref{tab:results} in
Sec.~\ref{sec:results}. There, the discretization was held fixed at $N=16$ and
the trial mass $M/m$ was varied to map out the bound-state spectrum. Here, by
contrast, we hold the physical point fixed at $M/m=1.6$ and examine how the
results change when the discretization is refined from $N=16$ to $32$, $64$, and $128$.
Table~\ref{tab:scaling} summarizes the fixed-mass results for these values of $N$. 
Its purpose is twofold: first, to show the convergence of the
physical eigenvalue and coupling with increasing quadrature resolution; second,
to collect the associated quantum-information diagnostics at the same physical
point, namely the entanglement entropy, Pauli-term count, and bond dimension
needed for high-fidelity MPS compression. At $M/m=1.6$, the minimum coupling
changes from $g^2_{\min}=479.97$ at $N=16$ to $493.5$ at $N=128$, i.e.
by only $2.8\%$, while $\lambda_{\max}$ changes by the same relative amount in
the opposite direction. The VQE proxy reproduces the exact variational optimum
to machine precision within the classical solver/backend
($<2\times10^{-13}$ relative error) at all five sizes ($N=8,16,32,64,128$); this number reflects numerical
optimizer agreement rather than the physical discretization uncertainty, which is
set by the $N$-dependence summarized in Tables~\ref{tab:scaling}. In the table, we further 
include the Richardson extrapolation to $N\to\infty$ using the $N=32,64$ pair, which gives
$g^2_{\min}(N{\to}\infty) \approx 494.0$ and $\alpha_\infty = g^2_\infty/(16\pi^2) \approx 3.128$. 
The $N=128$ point, sits 
close to this extrapolated limit, providing a nontrivial validation of the convergence trend.  
This analysis supports the
stability of the fixed-mass convergence trend without elevating it to a proof of
the full asymptotic continuum limit.

\begin{table}[tbp]
  \caption{BSE results at $M/m=1.6$, $\mu/m=0.15$ for five discretization sizes.
    $n$ = qubits, $N_{\rm Pauli}$ = non-zero Pauli terms,
    $\chi_{99\%}$ = bond dimension for $\geq 99\%$ MPS fidelity. Richardson's extrapolation uses the $N=32,64$ pair.}
  \label{tab:scaling}
  \begin{tabular}{cccccccc}
    \toprule
    $N$ & $n$ & $\lambda_{\max}$ & $g^2_{\min}$ & $\alpha = g^2/(16\pi^2)$ & $S_{\rm vN}^{\rm max}$ &
    $N_{\rm Pauli}$ & $\chi_{99\%}$ \\
    \midrule
    8   & 3 & $1.9590\times10^{-3}$ & 510.48& 3.233 & 0.178 & 36  & 1 \\
    16  & 4 & $2.0835\times10^{-3}$ & 479.97 & 3.039 & 0.568 & 136  & 2 \\
    32  & 5 & $2.0330\times10^{-3}$ & 491.88 & 3.115 & 0.566 & 528  & 2 \\
    64  & 6 & $2.0264\times10^{-3}$ & 493.48 & 3.125 & 0.554 & 2080 & 2 \\
    128 & 7 & $2.0261\times10^{-3}$ & 493.55 & 3.125 & 0.544 & 8256 & 2 \\
    $\infty$ (Richardson)&--&--& $494.0\pm0.5$ & $3.128\pm0.003$ &--& --& --\\
    \bottomrule
  \end{tabular}
\end{table}

\begin{figure}[tbp]
  \includegraphics[width=1.0\textwidth]{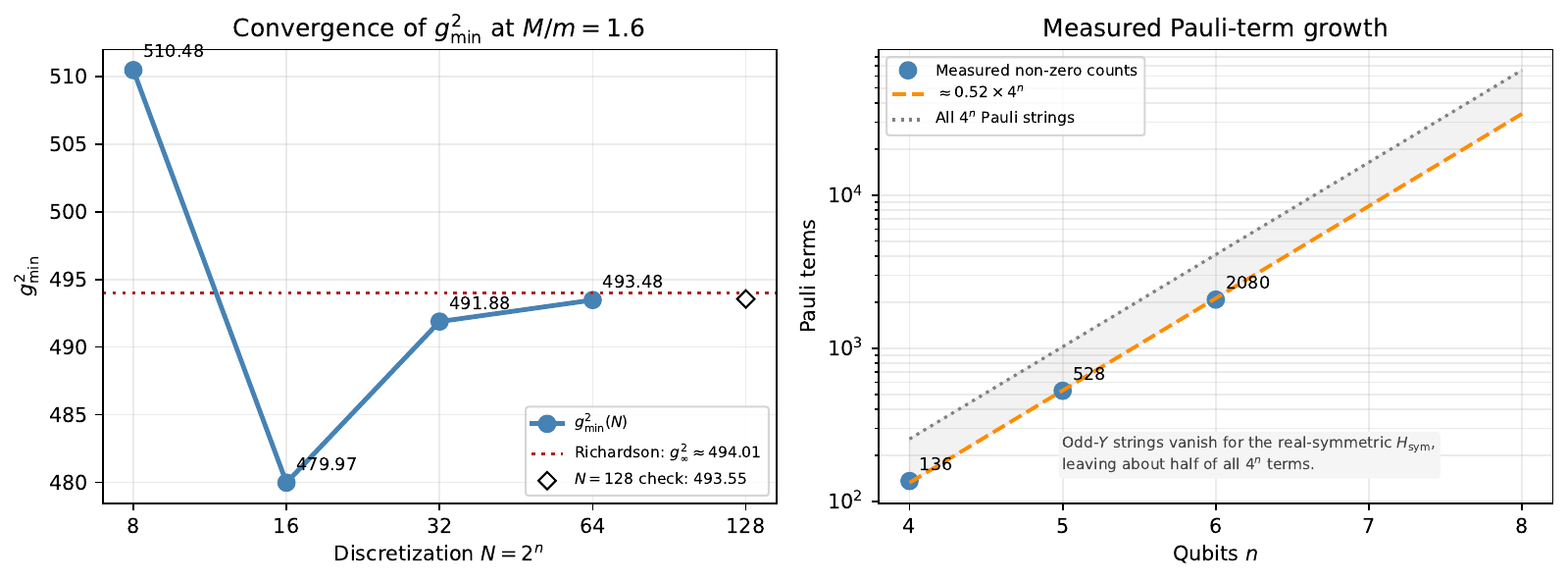}
  \caption{(Left) Convergence of $g^2_{\min}$ with $N$, with power-law fit and
    Richardson extrapolation to $g^2_\infty \approx 494.0$, corresponding to
    $\alpha_\infty = g^2_\infty/(16\pi^2) \approx 3.128$. The open diamond marks
    classical $N=128$ validation point, $g^2_{\min}=493.55$,
    which sits close to the extrapolated limit using only the $N=32,64$ pair.
    (Right) Pauli term count: non-zero terms follow the exact formula $(4^n+2^n)/2$
    (prefactor $0.531$ at $n=4$ decreasing to $0.504$ at $n=7$, approximately
    $0.52\times4^n$ over the measured range), because the real-symmetric $H_{\rm sym}$ has
    vanishing coefficients for all Pauli strings with an odd number of $Y$ factors.}
  \label{fig:richardson_pauli}
\end{figure}

The left panel of Fig.~\ref{fig:richardson_pauli}
visualizes this sequence, shows that the $N=16$ point already lies close to
the asymptotic regime, and includes the independent $N=128$ validation marker.
Richardson's extrapolation on
the $N=32,64$ pair, assuming $\mathcal{O}(1/N^2)$ convergence gives
$g^2_{\min}(N{\to}\infty) \approx 494.0$, i.e., $\alpha_\infty = 3.128$.
The quoted uncertainty $\pm 0.5$ is a conservative empirical estimate obtained
as the larger of (a) the residual
$|R(32,64) - g^2(128)| = 0.46$ and (b) the spread among
$R(32,64)$, $R(16,32)$, and $R(64,128)$ pairs.

\subsection{\texorpdfstring{Area-law entanglement as a function of $N$}{Area-law entanglement as a function of N}}

The main observation is that the maximum bipartite entropy remains nearly
independent of system size over the range studied:
$S_{\rm vN}^{\rm max} = 0.568$, $0.566$, and $0.554$ bits for
$N=16$, $32$, and $64$, respectively. This is far below the maximal entropy of
the central cut, which grows as $n/2$ bits for an $n$-qubit register. Thus, in
the present qubit encoding, the BSE amplitude is consistent with area-law
behavior with a size-independent effective bond dimension. The additional
classical $N=128$ check gives $S_{\rm vN}^{\rm max}=0.5439$ with
$\chi_{99\%}=2$, which is consistent with the same low-entanglement picture.

\begin{figure}[tbp]
 
  \includegraphics[width=1.0\textwidth]{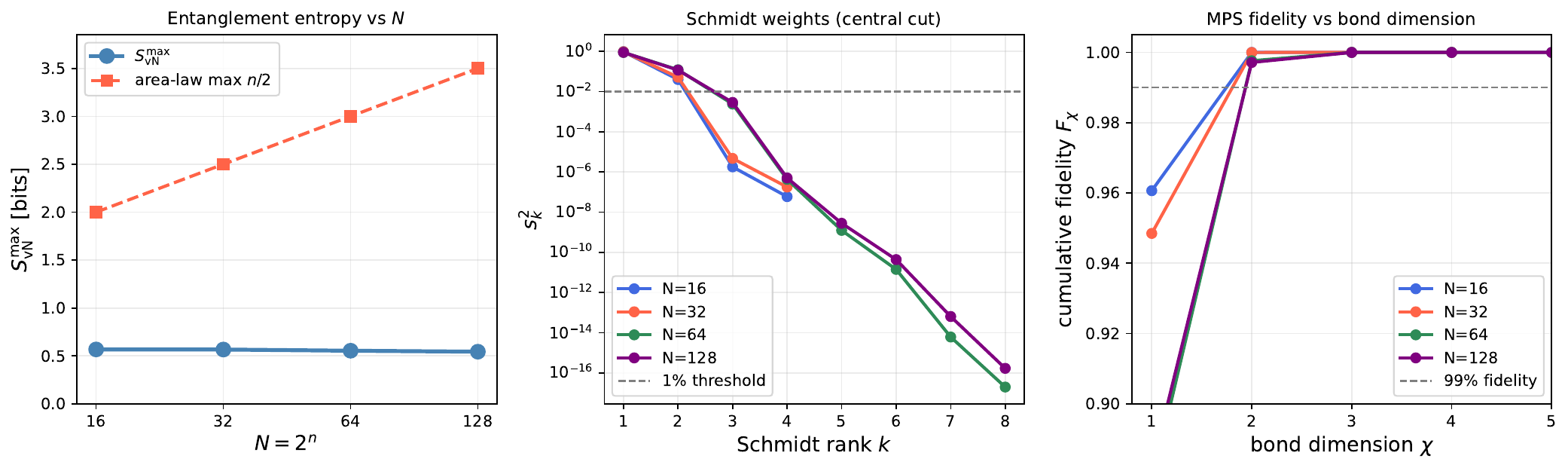}
  \caption{Entanglement analysis for $N=16$, $32$, $64$.
    (Left) $S_{\rm vN}^{\rm max}$ is flat in $N$, far below the area-law maximum.
    (Center) Schmidt weights $s_k^2$ decay steeply, almost identically for all sizes.
    (Right) Cumulative MPS fidelity: $\chi=2$ achieves $\geq 99\%$ at all $N$.}
  \label{fig:entanglement_scaling}
\end{figure}

The practical consequence is that the MPS compression cost does not grow with
$N$ over the tested range: Table~\ref{tab:scaling} shows that
$\chi_{99\%}=2$ for all three discretizations. Figure~\ref{fig:entanglement_scaling}
shows that the Schmidt spectra are likewise very similar across $N$. For the
largest case, $N=64$, the central-cut weights are
$s_1^2 = 87.7\%$, $s_2^2 = 12.0\%$, and $s_3^2 = 0.24\%$, so even
$\chi=3$ already gives $99.99\%$ fidelity.

\subsection{Explicit DMRG benchmark}

To demonstrate that the low-entanglement structure can indeed be exploited by
classical tensor-network methods, we perform explicit DMRG-2 calculations
using the \texttt{quimb} library~\cite{quimb2018}. Table~\ref{tab:dmrg}
shows the recovered $\lambda_{\max}$ and relative error for bond dimensions
$\chi\in\{2,4,8\}$ across $N=16$, $32$, and $64$, at the fixed benchmark
point $M/m=1.6$, $\mu/m=0.15$.

\begin{table}[tbp]
  \caption{DMRG-2 benchmark at $M/m=1.6$, $\mu/m=0.15$: recovered
    $\lambda_{\max}$ and relative error versus exact diagonalization for
    bond dimensions $\chi=2,4,8$. All runs use 20 sweeps with SVD cutoff
    $10^{-10}$.}
  \label{tab:dmrg}
  \begin{tabular}{cccc}
    \toprule
    $\chi$ & $N=16$ & $N=32$ & $N=64$ \\
    \midrule
    2 & $< 3\times10^{-3}$ & $< 3\times10^{-3}$ & $< 3\times10^{-3}$ \\
    4 & $< 3\times10^{-9}$ & $< 3\times10^{-9}$ & $< 3\times10^{-9}$ \\
    8 & $< 3\times10^{-9}$ & $< 3\times10^{-9}$ & $< 3\times10^{-9}$ \\
    \bottomrule
  \end{tabular}
\end{table}

Even at the minimal bond dimension $\chi=2$ the DMRG algorithm reproduces
$\lambda_{\max}$ to better than $3\times10^{-3}$ relative error for all
tested sizes — note that this reflects the algorithmic convergence of the
DMRG sweeps rather than the theoretical Schmidt truncation bound, which is
tighter (Sec.~\ref{sec:results}). Increasing to $\chi=4$ drives the error
below $3\times10^{-9}$, and $\chi=8$ yields no further improvement,
confirming that the BSE ground state in this encoding is efficiently captured
by low-rank MPS without any quantum resources.

\subsection{VQE convergence and Pauli scaling}

\begin{figure}[tbp]

  \includegraphics[width=1.0\textwidth]{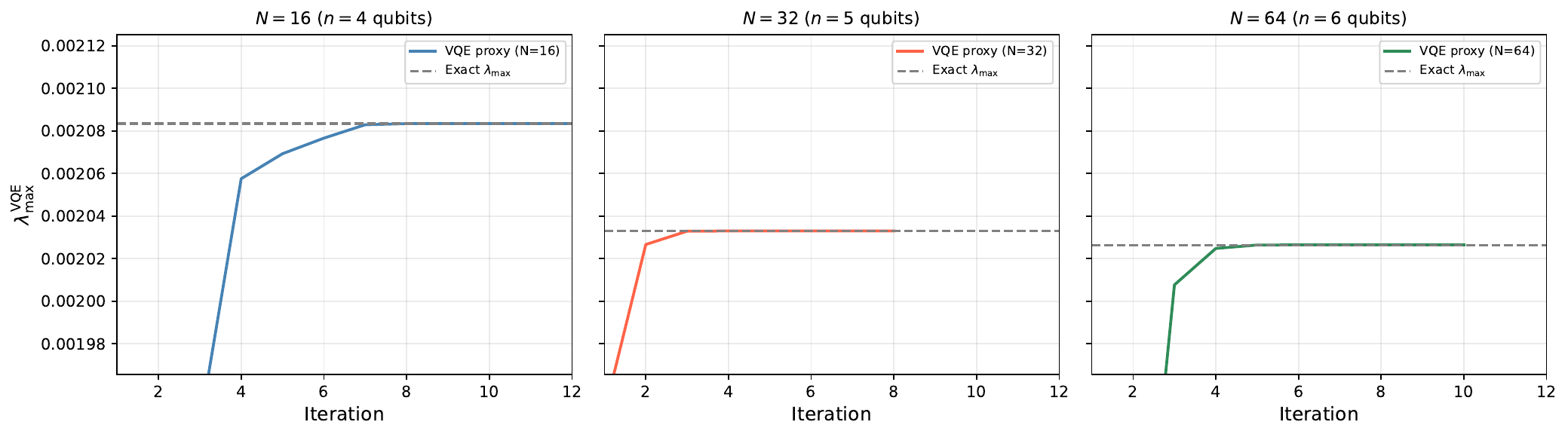}
  \caption{VQE convergence for $N=16$ (left), $N=32$ (center), $N=64$ (right)
    at $M/m=1.6$, all converging to exact $\lambda_{\max}$ (dashed) with
    essentially the same number of iterations.}
  \label{fig:vqe_scaling}
\end{figure}

Figure~\ref{fig:vqe_scaling} shows that the variational optimization remains
qualitatively similar for $N=16$, $32$, and $64$: in all three cases the VQE
proxy converges to the exact $\lambda_{\max}$ within essentially the same
iteration range. Over this size window, increasing $N$ therefore does not by
itself make the optimization visibly harder.

The Hamiltonian representation, however, does grow rapidly with qubit number.
As shown in Table~\ref{tab:scaling} and in the right panel of
Fig.~\ref{fig:richardson_pauli}, the number of nonzero Pauli coefficients is
given exactly by $(4^n + 2^n)/2$, which gives 136, 528, 2080, and 8256 for
$n=4,5,6,7$ respectively. As a fraction of all $4^n$ Pauli strings, this equals
$0.5312$ at $n=4$, $0.5156$ at $n=5$, $0.5078$ at $n=6$, and converges to
$0.5$ as $n\to\infty$; the figure caption quotes $\approx 0.52\times 4^n$ as a
convenient round figure over the measured range.
The sub-half-filling reflects the real-symmetric structure of $H_{{\rm sym}}$, which
forces all Pauli strings with an \emph{odd} number of $Y$ operators to have
zero coefficient.

\section{\label{sec:outlook}Analysis of the findings and Quantum Advantage}

\subsection{Three independent barriers}

We now give a careful and self-critical assessment of whether extending the present
approach to large qubit numbers ($n \gtrsim 100$) would yield a quantum advantage.
For the \emph{specific problem studied here}, namely the Euclidean S-wave BSE in the
ladder approximation, the answer is \textbf{no}, for three independent reasons that each
individually suffice to prevent a speedup. This negative result
characterizes the present problem and identifies which extensions
\emph{may} become substantially harder. 

\textbf{Barrier 1: Pauli overhead.}
The Pauli decomposition of $H_{\rm sym}$ contains $(4^n+2^n)/2$ non-zero
terms exactly (Table~\ref{tab:scaling}), ranging from a prefactor of $0.531$
at $n=4$ to $0.504$ at $n=7$, asymptotically approaching $\tfrac{1}{2}\times 4^n$.
For the tensor-network calculations reported
here, $\langle \hat H_{\vqe} \rangle$ is evaluated exactly and no sampling cost is
incurred. On actual gate hardware, however, the energy of a generic dense
Hermitian operator is not available from a single native measurement: one needs
either Hamiltonian averaging over a Pauli decomposition, measurements of
commuting groups, or a more structured oracle-based construction. In the
standard VQE setting, each estimated expectation value carries shot noise, with
$\mathcal{O}(\varepsilon^{-2})$ samples needed to reach additive precision
$\varepsilon$~\cite{McClean2016,Tilly2022}. Therefore, for the present explicit
Pauli representation, any straightforward sampling-based implementation still
inherits a cost tied to the exponentially large operator description. For
$n=100$, one would have
$N_{\rm Pauli} \approx \tfrac{1}{2}\times 4^{100} \approx 8\times 10^{59}$ nonzero terms,
which is already prohibitive at the level of operator storage and naive
measurement scheduling. This does not exclude more advanced grouping,
shadow-based estimation, or fault-tolerant block-encoding methods; it shows
that the explicit Pauli-sum VQE strategy used here does not scale favorably to
large $n$.

\begin{figure}[tbp]

  \includegraphics[width=1.0\textwidth]{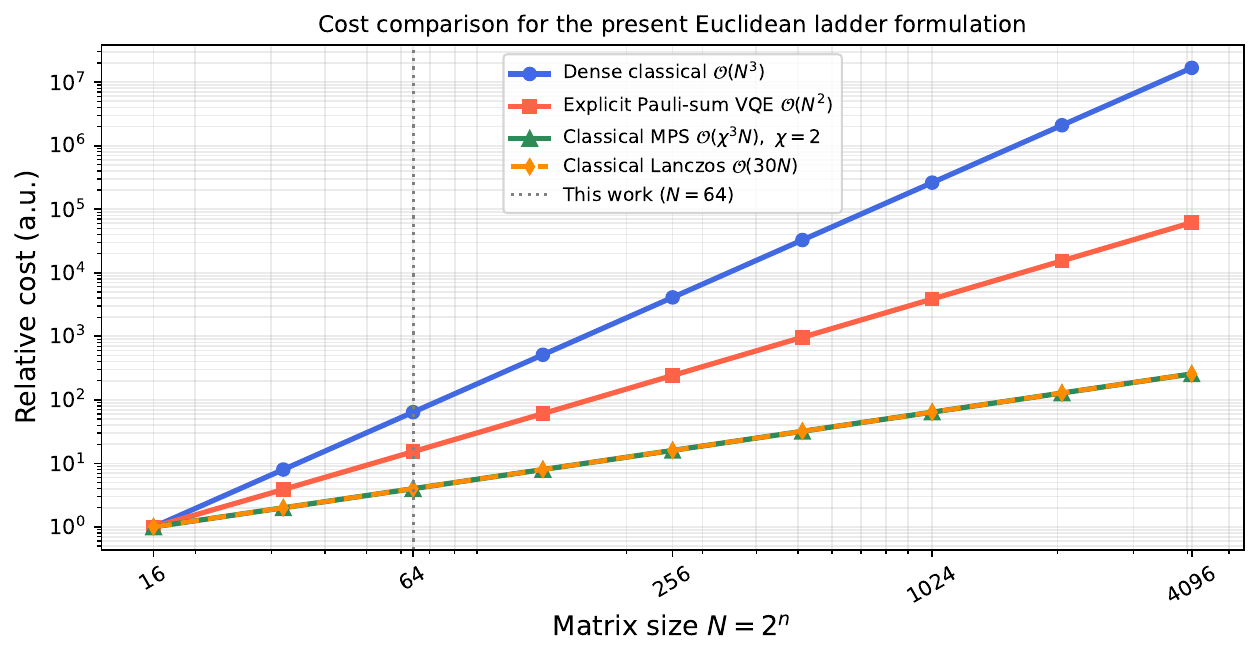}
  \caption{Scaling of the different methods to solve the ladder BSE: 
  Dense Classical,
  Classical Lanczos, VQE with explicit Pauli decomposition, 
  and classical MPS-based methods. Note, the lines for 
  classical Lanczos and classical MPS-based methods 
  coincide in the figure.}
  \label{fig:quantum_advantage}
\end{figure}

\textbf{Barrier 2: Area law — the decisive argument.}
Our central finding is that, over the range studied, the BSE amplitude shows
area-law-like entanglement in the qubit encoding (Sec.~\ref{sec:results},
Fig.~\ref{fig:entanglement_scaling}):
\begin{equation}
  S_{\rm vN}^{(\rm max)}(N) \in [0.54,\,0.57]~\text{bits} \ll \frac{n}{2}~\text{bits}
  \quad \text{for all } N \in \{16,32,64,128\},
  \label{eq:area_law_const}
\end{equation}
with a weak decreasing trend ($0.568$ at $N=16$ to $0.544$ at $N=128$),
while the theoretical area-law ceiling $n/2$ bits grows linearly with $n$.
This is precisely the regime in which classical tensor-network methods are most
effective. Empirically, the Schmidt spectra in Sec.~\ref{sec:largeN} decay so
rapidly that $\chi_{99\%}=2$ for $N=16$, $32$, $64$, and $128$, meaning that the
numerically relevant state is already captured by a very small bond dimension.
For fixed small $\chi$, a classical MPS or DMRG-type treatment scales only
linearly in the system size,
\begin{equation}
  C_{\rm MPS} = \mathcal{O}(\chi^3 \cdot N) = \mathcal{O}(8N),
\end{equation}
for $\chi=2$~\cite{White1992,Schollwoeck2011,Vidal2003}.
This is confirmed by the explicit DMRG benchmarks in Table~\ref{tab:dmrg},
which show that $\chi=2$ suffices for $<3\times10^{-3}$ relative error at all tested~$N$.
A classical Lanczos iteration for the maximum eigenvalue
performs $k\approx30$ matrix--vector
products to converge. For the dense angular-averaged kernel each product costs
$\mathcal{O}(N^2)$, giving $\mathcal{O}(kN^2)=\mathcal{O}(30N^2)$; if a fast
$\mathcal{O}(N)$ matrix--vector product is available the cost reduces to
$\mathcal{O}(kN)=\mathcal{O}(30N)$ (the value plotted as
``$\mathcal{O}(30N)$'' in Fig.~\ref{fig:quantum_advantage}).
Within the low-entanglement regime observed here, both approaches remain far
more efficient than the explicit Pauli-sum VQE realization studied in this work.

The key point is therefore empirical rather than theorem-level: across the
available sizes $N = 16, 32, 64$, the Schmidt weights at the central
bipartition decay as $s_1^2 \approx 0.88$--$0.96$, $s_2^2 \approx 0.04$--$0.12$,
and $s_{k\geq3}^2 < 10^{-3}$. This is exactly the pattern expected for a state
that is efficiently compressible as an MPS. It strongly suggests that the
Euclidean O(4) S-wave problem remains classically tractable, although it does
not by itself constitute a rigorous asymptotic proof for all larger $N$.
Figure~\ref{fig:quantum_advantage} summarizes the cost of the methods as a
function of the matrix size $N=2^n$ on a log--log scale. Because $N=2^n$, all
four methods scale exponentially in the qubit number $n$; what differs is the
\emph{power of $N$}: dense diagonalization is $\mathcal{O}(N^3)$, the explicit
Pauli-sum VQE proxy is $\mathcal{O}(N^2)$ (the Pauli-term count grows as
$\tfrac12 4^n=\tfrac12 N^2$), and the classical Lanczos and low-bond-dimension
MPS contractions are both $\mathcal{O}(N)$ and therefore coincide on the plot.
The advantage of the MPS/Lanczos methods here is thus a lower polynomial degree
in $N$, enabled by the area-law structure --- not a change from exponential to
linear scaling in $n$.

It is worth stating precisely \emph{why} the entanglement is low, since the
matrices $K_{\rm sym}$ and $H_{\rm sym}$ are in fact dense. 
The low tensor-network entanglement (in the quantics bit-partition of the index)
is a property of the dominant \emph{eigenvector}, not of the density or
conditioning of $H_{\rm sym}$: under the binary index map
$i=(b_1\ldots b_n)$ the amplitude $\phi(p_i)$ becomes a function sampled on a
dyadic grid, and smooth single-scale functions of one variable have rapidly
decaying Schmidt coefficients across the bit-partitions of this ``quantics''
encoding, hence admit low-bond-dimension (quantized tensor-train)
representations~\cite{Oseledets2011,Khoromskij2011}. The BSE S-wave amplitude is
exactly such a function --- a single smooth peak in $p$
(Fig.~\ref{fig:amplitude}) --- which is why $\chi\approx2$ suffices. Equivalently,
the relevant ``spectral'' statement is that the reduced density matrices
$\rho_k$ have a rapidly decaying spectrum (the Schmidt weights $s_a^2$ plotted
above); it is the spectrum of $\rho_k$, not of $H_{\rm sym}$, that controls
compressibility.

A useful baseline is what a \emph{generic} problem would look like. The
eigenvectors of the Gaussian ensembles GOE/GUE/GSE are Haar-random in the
real/complex/quaternionic invariant ensembles, and a Haar-random eigenvector of
a $2^n$-dimensional space is, with overwhelming probability, near-maximally
entangled: by Page's theorem the average bipartite entropy of a balanced cut is
$\langle S\rangle\approx \tfrac{n}{2}-\tfrac{1}{2\ln2}$
bits~\cite{Page1993}, with a broad Marchenko--Pastur Schmidt
spectrum~\cite{ZyczkowskiSommers2001} and Schmidt rank of order $2^{n/2}$. Such
a state would require an \emph{exponentially} large bond dimension
$\chi\sim2^{n/2}$ for a faithful MPS. The BSE ground state sits at the opposite
extreme: its measured $S_{\rm vN}^{\max}\in[0.54,0.57]$ bits is essentially
independent of $n$ and lies far below the Page ceiling $n/2$, so $\chi=2$
suffices. This contrast also answers which problems are promising for VQE/quantum
methods: not smooth, low-entanglement targets like the present one, but those
whose ground states carry genuinely high, structured (area-law-in-2D or
volume-law) entanglement --- precisely the BSE extensions collected in
Eq.~\eqref{eq:hierarchy}.

\begin{figure}[tbp]
  \includegraphics[width=1.0\textwidth]{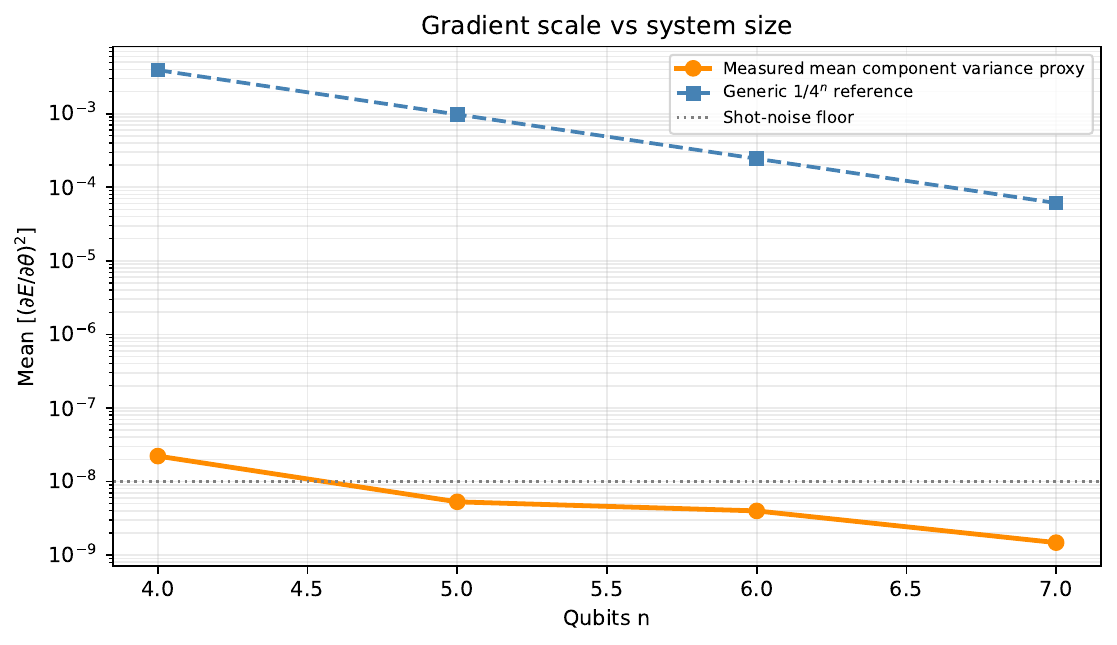}
  \caption{Gradient diagnostics for the MPS-style VQE ansatz.
    Mean squared gradient
    $\langle(\partial E/\partial\theta)^2\rangle$ vs qubit number $n$:
    measured values (circles) decline more slowly than the generic $1/4^n$
    barren-plateau reference (dashed). The gray line marks an indicative
    shot-noise floor.}
  \label{fig:gradient_stats}
\end{figure}

\textbf{Barrier 3: Barren plateaus.}
For shallow hardware-efficient ansätze, the gradient variance of the VQE loss
function decays as~\cite{McClean2018barren}:
\begin{equation}
  \mathrm{Var}\!\left[\frac{\partial \langle H \rangle}{\partial \theta_k}\right]
  \sim \frac{1}{4^n}.
  \label{eq:barren}
\end{equation}
If this generic scaling applies, then by $n=15$ one expects a variance of order
$10^{-9}$, making gradients increasingly difficult to resolve in the presence of
sampling noise and finite numerical precision. This does not imply a sharp
failure threshold, nor does it prove that the specific ansatz used here must
develop a barren plateau at that exact qubit number. It does show, however,
that gradient-based VQE training becomes progressively less reliable as $n$
grows unless additional structure is exploited. In the companion calculations for
the present MPS-style ansatz, the mean squared gradient decreases from
$2.2\times 10^{-8}$ at $n=4$ to $4.0\times 10^{-9}$ at $n=6$
(and to $1.5\times 10^{-9}$ at $n=7$), showing a clear
degradation trend over the tested range, although not a clean $1/4^n$ law.
Figure~\ref{fig:gradient_stats} visualizes this decay together with the
generic $1/4^n$ reference line and gradient-magnitude diagnostics.
Barren-plateau mitigation
strategies such as layerwise training or problem-informed ansätze can help, but
they do not remove the general large-$n$ risk~\cite{Cerezo2021}.

The three barriers together do not amount to a formal complexity-class theorem
for the Euclidean S-wave BSE. What they do show is that, for the present
O(4)-symmetric ladder problem and over the tested range $N\leq 64$, the state
remains only weakly entangled and is therefore highly compressible. Combined
with the known efficient classical simulation of slightly entangled and
matrix-product-state-representable systems~\cite{Vidal2003,White1992,Schollwoeck2011,Eisert2010area},
where Vidal gives the quantum-information statement, White and Schollw\"ock
establish the practical DMRG/MPS setting, and Eisert \emph{et al.} review the
area-law--simulability connection,
this provides strong evidence that the present problem lies in a classically
tractable regime and is unlikely to exhibit end-to-end quantum advantage in
its current explicit-Pauli-sum/VQE formulation. In that sense, the Euclidean
ladder BSE is closer to one-dimensional problems efficiently treated by DMRG
than to the harder targets that usually motivate quantum algorithms.

\subsection{Where quantum advantage becomes plausible}

The same analysis identifies which \emph{extensions} of the BSE may leave this
classically tractable regime. Three physically motivated scenarios are expected
to produce substantially stronger entanglement and nonlocal couplings, making
classical tensor-network methods less effective and quantum advantage more
plausible.

\textbf{(i) 2D Minkowski-space BSE.}
A direct Minkowski-space solution (avoiding the Nakanishi detour) requires discretizing
\emph{two} independent momentum variables $(p^+, p^\perp)$ in light-front coordinates.
The resulting 2D integral kernel couples distant momentum points,
generating long-range correlations. By the area-law theorem for 2D systems, the
entanglement entropy of a region of linear size $n_s$ qubits (with total register
size $n_{\rm tot}=n_s^2$) scales as $S \sim n_s \sim \sqrt{n_{\rm tot}}$~\cite{Eisert2010area},
rather than being bounded by a constant as seen here, and the classical MPS bond
dimension grows exponentially with $n_s$. For $n_s = 10$
qubits per spatial dimension, the problem requires a $100$-qubit register where
classical MPS becomes intractable.

\textbf{(ii) Three-body BSE and Faddeev equations.}
The relativistic three-body bound state requires a Faddeev-type kernel coupling
three relative momenta simultaneously \cite{Eichmann2016}. The resulting tensor-product Hilbert space
has $3n$ qubits and the interaction graph is non-planar, breaking the 1D
topology that underlies the area law. Volume-law entanglement is expected,
and quantum algorithms based on Quantum Phase Estimation
(QPE)~\cite{Cerezo2021,Kandala2017} — which avoid barren plateaus entirely —
are the natural approach on fault-tolerant hardware.

\textbf{(iii) Non-ladder kernels in QCD.}
Beyond the ladder approximation, crossed-box and self-energy diagrams dress
the propagators and generate effective long-range interactions in momentum space.
This is also the regime in which the limitations of the ladder approximation
itself become more pronounced in Bethe-Salpeter spectral studies~\cite{AhligAlkofer1999}.
Dyson-Schwinger-based dressed kernels~\cite{Roberts1994,Barabanov2021,Ding2024} are known to produce
large confining quark-gluon correlations. Whether these produce volume-law
entanglement in the BSE qubit encoding is an open question that can now be
addressed using the framework developed in this paper.

The hierarchy is:
\begin{equation}
  \begin{aligned}
    &\underbrace{\text{Euclidean S-wave ladder}}_{\text{classically tractable regime}}
    \;\subset\;
    \underbrace{\text{Minkowski / 2D / crossed-box}}_{\text{plausible quantum-advantage targets}} \\
    &\;\subset\;
    \underbrace{N\text{-body QCD}}_{\text{likely fault-tolerant targets}}
  \end{aligned}
  \label{eq:hierarchy}
\end{equation}

Our result demonstrating \emph{absence} of quantum advantage provides 
the first \emph{entanglement quantification} of the BSE amplitude
in qubit encoding, to our knowledge, showing $S_{\rm vN}^{\rm max}$
in the range $0.54$--$0.57$ bits across $N=16$--$128$, well below the
area-law ceiling and with a weak decreasing trend.
In addition, it establishes the BSE Pauli Hamiltonian framework used here 
as a \emph{diagnostic tool}: applying the same entanglement analysis to extended
BSE variants (Minkowski, $N$-body, non-ladder) directly measures whether
those variants move beyond the low-entanglement regime seen here, without
requiring a full quantum simulation.
Furthermore, it identifies the O(4) S-wave projection as the specific symmetry responsible
for the area law. Breaking this symmetry — by moving to Minkowski space,
adding partial waves, or using non-ladder kernels — is the precise prescription
for reaching the quantum-advantage regime.

\section{\label{sec:conclusions}Conclusions}

We have presented, to our knowledge, the first gate-based quantum computing solution of the
homogeneous Bethe-Salpeter equation for relativistic scalar bound states,
using VQE with a Matrix Product State tensor-network ansatz.

The main results established in this work are:
\begin{enumerate}
  \item The hBSE reduces to a symmetric eigenvalue problem for $N=2^n$
        discretization points, directly amenable to VQE on $n$ qubits.
  \item The BSE amplitude shows area-law-like entanglement in the qubit encoding:
      $S_{\rm vN}^{\rm max} = 0.568$, $0.566$, $0.554$, $0.544$ bits for
      $N=16$, $32$, $64$, $128$ respectively — far below the area-law ceiling,
      with a weak decreasing trend consistent with the area law becoming
      more effective at larger~$N$,
        and bond dimension $\chi=2$ gives $\geq 99\%$ MPS fidelity at all tested sizes.
  \item For $n=4,5,6,7$ qubits ($N=16,32,64,128$), the L-BFGS-B VQE proxy
        (classical noiseless optimizer applied to the same variational objective)
        reproduces classical diagonalization to better than $10^{-13}$ relative error.
        This is a numerical-precision statement about the classical optimizer, not a
      physical uncertainty estimate, and is distinct from the ADAM-based PennyLane
      VQE, which achieves $<1\%$ mean relative error (best configuration:
      $\bar\delta_{\rm rel}=4.8\times10^{-3}$, Table~\ref{tab:vqe_grid_summary}).
      The minimum
        coupling $g^2_{\min}$ converges to within $2.8\%$ between $N=16$ and $N=64$.
      An independent classical $N=128$ check gives $g^2_{\min}=493.55$,
      only $0.015\%$ above the $N=64$ value and within $0.093\%$ of the
      Richardson estimate.
    \item A critical analysis of three independent barriers (Pauli overhead $\sim 4^n$,
        area-law entanglement $S_{\rm vN}\in[0.54,0.57]$ bits — slowly decreasing and
        well below the area-law ceiling — and a measured
        decrease in gradient scale together with the generic barren-plateau literature)
      shows that the present problem — Euclidean S-wave
      ladder BSE — lies in a classically tractable regime and is well handled by
      MPS/Lanczos algorithms over the tested range. No end-to-end quantum
      advantage is expected for the present explicit-Pauli-sum/VQE formulation.
\end{enumerate}
This conclusion does not diminish the contribution: the entanglement
quantification ($S_{\rm vN}^{\rm max}$ ranging from $0.568$ bits at $N=16$
to $0.544$ bits at $N=128$, consistently far below the area-law ceiling) is a new
result, the Pauli Hamiltonian encoding is a reusable framework, and the
analysis identifies which BSE extensions are the most credible candidates for
future quantum advantage studies.

The hierarchy Eq.~\eqref{eq:hierarchy} should therefore be read as a roadmap,
not as a demonstrated quantum speedup claim. The most plausible future targets
are (i) 2D Minkowski-space BSE with two momentum variables, which can break the
low-entanglement structure seen here; (ii) three-body Faddeev-BSE systems with
non-planar interaction graphs and expected volume-law entanglement; (iii)
non-ladder kernels from Dyson-Schwinger equations~\cite{Roberts1994}; and (iv)
fault-tolerant implementations based on Quantum Phase Estimation, which avoid
variational barren-plateau issues entirely. These directions are physically
motivated extrapolations beyond the present evidence, rather than results
already established in this paper.

\backmatter

\bmhead{Acknowledgements}
We thank Gernot Eichmann for useful discussions about recent developments on Bethe-Salpeter equations.

\section*{Declarations}

\begin{itemize}
\item Funding: This work was supported by the Ministry of Economic Affairs, Labour and Tourism Baden-Württemberg in the frame
of the Competence Center Quantum Computing Baden-Württemberg (Projects No. KQCBW25) and by the 
German Federal Ministry of Research, Technology and Space within the funding program
‘Application-oriented quantum computing’ under Contract No. 13N17159.
\item Conflict of interest/Competing interests: The authors declare that they have no conflict of interest.
\item Ethics approval and consent to participate: Not applicable.
\item Consent for publication: Not applicable.
\item Data availability: Not applicable.
\item Materials availability: Not applicable.
\item Code availability: The code that supports the findings of this study is available from the corresponding author upon reasonable request.
\item Author contribution: Not applicable.
\item 
\end{itemize}

\appendix

\section{\label{app:kernel}Derivation of the O(4) S-wave Kernel}

The exchange kernel in Euclidean 4D reads $1/[(p_E - q_E)^2 + \mu^2]$.
We project onto the O(4) S-wave by integrating over the solid angle $S^3$
with the normalized surface element $(2\pi^2)^{-1}d\Omega_3$
(see Ref.~\cite{NieuwenhuisTjon1996} for a detailed derivation):
\begin{equation}
  \KO(p,q;\mu) = \frac{1}{2\pi^2}\int_{S^3}\frac{d\Omega_3}{(p_E-q_E)^2+\mu^2}.
\end{equation}
Writing $(p_E-q_E)^2 = p^2 + q^2 - 2pq\cos\theta$ and using the $S^3$
surface element $d\Omega_3 = 4\pi\sin^2\theta\,d\theta$, where
$\theta$ is the polar angle relative to the fixed external momentum and the
remaining $S^2$ has been integrated out ($\int d\Omega_2 = 4\pi$):
\begin{equation}
  \KO = \frac{2}{\pi}\int_0^\pi\frac{\sin^2\theta\,d\theta}{p^2+q^2-2pq\cos\theta+\mu^2}.
\end{equation}
Setting $A = p^2+q^2+\mu^2$, $B = 2pq$, and using
\newline
$\int_{-1}^{1}(1-x^2)/(A-Bx)\,dx = (B^2-A^2)\ln\!\bigl[(A{+}B)/(A{-}B)\bigr]/B^3 + 2A/B^2$,
\newline
one can verify (numerically: $|\KO^{\rm alg} - \KO^{\rm int}|/\KO^{\rm int} < 10^{-15}$)
that the integral equals:
\begin{equation}
  \KO(p,q;\mu) = \frac{A - \sqrt{A^2-B^2}}{2p^2 q^2},
  \quad A^2-B^2 = \bigl[(p{-}q)^2{+}\mu^2\bigr]\bigl[(p{+}q)^2{+}\mu^2\bigr],
\end{equation}
which is Eq.~\eqref{eq:kernel_O4}. The equality is an algebraic identity
verified analytically and confirmed numerically.

\section{\label{app:pennylane}Implementation details}

Key features:
\begin{itemize}
  \item Device: \texttt{qml.device("default.tensor", method="mps", max\_bond\_dim=e.g. 8)}
  \item Ansatz: \texttt{qml.MPS} template with RY + CNOT blocks
  \item Optimizer: ADAM, step size 0.03, tolerance $10^{-8}$
  \item Gradient: parameter-shift rule (exact, device-independent)
  \item Construction of the symmetrized BSE Hamiltonian $H_{\rm sym}$ 
  for arbitrary $N=2^n$ using Gauss-Legendre quadrature and the O(4) 
  S-wave projected kernel.
  \item Exact classical diagonalization for benchmarking, 
  with all eigenvalues and eigenvectors computed for reference.
  \item Explicit Pauli decomposition of $H_{\rm sym}$ into $n$-qubit 
  Pauli strings, with the number of nonzero terms scaling 
  as $\sim 0.5 \times 4^n$ due to the real-symmetric structure.
  \item VQE implementation using a matrix product state (MPS) 
  ansatz with alternating RY rotations and CNOTs, optimized on 
  the PennyLane tensor-network backend (quimb).
  \item Grid sweep over ansatz depth and optimizer step size, 
  with statistical analysis over multiple random initializations.
  \item Entanglement analysis: calculation of the von Neumann entropy 
  and Schmidt weights for all bipartitions, confirming area-law scaling
  and rapid decay of the Schmidt spectrum.
  \item Scaling studies for $N=8,16,32,64,128$, including classical-only 
  validation at $N=128$ and explicit DMRG/MPS benchmarks.
\end{itemize}

The code is compatible with PennyLane $\geq$ 0.37.

\bibliographystyle{unsrt}
\bibliography{refs}

\end{document}